\documentclass[hyper,a4paper]{JHEP3}
\usepackage{multirow}
\pdfoutput=1
\usepackage{cite}
\usepackage{subfigure}
\usepackage{rotating}
\usepackage{amssymb}
\usepackage{amsmath}
\usepackage{blkarray}
\usepackage{multirow}
\usepackage{mathtools}
\usepackage{relsize}

\newcommand{\bea}{\begin{equation}}
\newcommand{\eea}{\end{equation}}
\newcommand{\be}{\begin{eqnarray}}
\newcommand{\ee}{\end{eqnarray}}

\def\hbar#1{\backslash\hspace{-2mm}#1}

\catcode`\@=11
\def\lsim{\mathrel{\mathpalette\@versim<}}
\def\gsim{\mathrel{\mathpalette\@versim>}}
\def\@versim#1#2{\vcenter{\offinterlineskip
\ialign{$\m@th#1\hfil##\hfil$\crcr#2\crcr\sim\crcr } }}
\catcode`\@=12

\def\2tvec#1#2{
\left(
\begin{array}{c}
#1  \\
#2  \\
\end{array}
\right)}

\def\mat2#1#2#3#4{
\left(
\begin{array}{cc}
#1 & #2 \\
#3 & #4 \\
\end{array}
\right) }

\def\Mat3#1#2#3#4#5#6#7#8#9{
\left(
\begin{array}{ccc}
#1 & #2 & #3 \\
#4 & #5 & #6 \\
#7 & #8 & #9 \\
\end{array}
\right) }

\def\3tvec#1#2#3{
\left( 
\begin{array}{c}
#1  \\
#2  \\
#3  \\
\end{array}
\right)}

\def\to{\rightarrow}

\def\ptmiss{\not\!\!{p_T}}

\def\hbar#1{\backslash\hspace{-2mm}#1}

\numberwithin{equation}{section}

\title{Multi-Lepton Signatures of the Triplet Like Charged Higgs at the LHC}

\author{Priyotosh Bandyopadhyay$^{a,b,1}$,  Katri Huitu$^{a,2}$ and Asl{\i} Sabanc{\i} Ke\c{c}eli$^{a,3}$\\

$^a$Department of Physics,  and Helsinki Institute of Physics,\\
P.O.Box 64 (Gustaf H\"allstr\"omin katu 2), FIN-00014 University of Helsinki, Finland\\
$^b$Dipartimento di Matematica e Fisica "Ennio De Giorgi", Universit`a del Salento and INFN, Via Arnesano, 73100, Lecce, Italy\\
Email: \email{$^1$priyotosh.bandyopadhyay@helsinki.fi, priyotosh.bandyopadhyay@le.infn.it,$^2$katri.huitu@helsinki.fi,
$^3$asli.sabanci@helsinki.fi}}

\abstract{We study multi-lepton signatures of the triplet like charged Higgs at the LHC in the context of $Y=0$ triplet extended supersymmetric model (TESSM).  
In TESSM the  $h_i^\pm W^\mp Z$  coupling appears at tree level when the triplet vacuum expectation value is nonzero, and because of the coupling the charged Higgs decay channels as well as the production channels can dramatically change at the LHC. 
We show that for the triplet dominated charged Higgs the main production channels are no longer through the top decay or $gg$ and $gb$ fusions since these are very suppressed due to the lack of triplet-SM fermion coupling. 
In the numerical analysis, we consider also other  possible production channels some of which have additional contributions from the diagrams containing $h_i^\pm W^\mp Z$ vertex.  
We investigate the decay channels of a triplet like light charged Higgs ($m_{h_1^{\pm}}\leq 200$ GeV) and show that depending on the triplet component, the charged Higgs can substantially decay to $W^\pm Z$.  
We further examine the $3l$, $4l$, $5l$ multi-lepton signatures of the triplet like charged Higgs by considering four different benchmark points for which we perform PYTHIA level simulation using FastJet for jet formation at the LHC with 14 TeV.  We found that for favorable parameters the earliest discovery with 5$\sigma$ signal significance can appear with early data of 72 fb$^{-1}$ of integrated luminosity. We also present the invariant mass distribution $M_{lljj}$ for ($\geq 3\ell)+(\ptmiss \geq 30$ GeV) and $(\geq 3\ell)+(\geq 2j)+(\ptmiss \geq 30$ GeV) and show that in addition to the charged Higgs mass peak, an edge that carries information about heavy intermediate neutral Higgs bosons arises at the end of the mass distribution.

}  

\preprint{HIP-2014-24/TH}

\begin{document}

\section{Introduction}
The discovery of the Higgs boson \cite{Higgsd1,Higgsd2} is a very important step toward understanding the nature of the electroweak symmetry breaking (EWSB) and its underlying model. Though the experimental results for Higgs production and decay channels are in very good agreement with the Standard Model (SM) predictions \cite{ZZ1,ATLAS:2013nma,WW1,ATLAS:2013wla,Chatrchyan:2013zna,ATLAS:2012aha,Chatrchyan:2014nva,tau2,CMS:ril,ATLAS:2013oma} there is still room for models beyond the SM which contain more than one Higgs boson in the particle spectrum. 
Such models are motivated by the problems of the SM such as unnaturalness of the Higgs mass, the lack of neutrino masses and a dark matter candidate.

The common feature of any extended Higgs sector possessing more than one Higgs doublet is to contain at least one charged Higgs pair along with the neutral Higgs bosons. 
Even though the discovery of a charged Higgs would be a clear evidence of a nonstandard Higgs model, the subsequent analysis of the properties of charged Higgs is needed to reveal the structure as well as the symmetries of the Higgs sector. 

The most studied possibility is to have only one charged Higgs boson pair ($h^{\pm}$) in the spectrum where the SM Higgs sector is extended by adding an extra Higgs doublet with or without supersymmetry (for reviews see {\it{e.g.}} Refs.\cite{Branco:2011iw} and \cite{Djouadi:2005gj}). In multi-doublet models, the charged Higgs can couple to the SM fermions and the main charged Higgs decay channels are $tb$ and/or $\tau \nu_{\tau}$ depending on the mass of $h^{\pm}$ \cite{Djouadi:1995gv,Djouadi:2005gj}. At the LHC, the light charged Higgs boson ($m_{h^{\pm}}\leq m_{t}-m_{b}$) is mainly produced by the top decay through $pp\to t\bar{t}$, $t\rightarrow h^{\pm} b$ \cite{Dittmaier:2012vm,Berger:2003sm} whereas  $gg\rightarrow tbh^{\pm}$ and $gb\rightarrow th^{\pm}$ channels become the dominant production channels for the heavy charged Higgs \cite{Gunion:1986pe,DiazCruz:1992gg,Barger:1993th,Plehn:2002vy,Weydert:2009vr}. 

Besides the models with several Higgs doublets, the charged Higgs can be originated from the models comprising  higher representations of Higgs fields such as triplets. 
In particular, the triplet models with non-zero triplet vacuum expectation value (vev) contains a tree level $h^{\pm}ZW^{\mp}$ coupling as a consequence of the custodial $SU(2)_c$ symmetry breaking. 
The appearance of this coupling at tree level is important for identifying the triplet's impact in the charged Higgs since $h^{\pm}ZW^{\mp}$ coupling is induced at loop-level in the models with only Higgs doublets \cite{Gunion:1987ke,Mendez:1990epa,CapdequiPeyranere:1990qk,Kanemura:1997ej}. Thus this vertex opens up a new decay channel $h^{\pm}\rightarrow ZW^{\mp}$ as well as the charged Higgs production channel through the vector boson fusion.  

To investigate the phenomenology of this coupling at LHC, we consider here the triplet extended supersymmetric model (TESSM) where the MSSM field content is extended by adding a $Y=0$ triplet chiral superfield. After the Higgs discovery, this model has become an attractive alternative to the MSSM. 
The reason for this is that  to accommodate a 125 GeV Higgs boson, in the MSSM large radiative corrections obtained via a large mixing between third generation squarks and/or very heavy stop masses \cite{Carena:2011aa} are needed. 
However, such a  heavy spectrum results in a severe fine-tuning $\lesssim 1\%$ \cite{CahillRowley:2012rv,CahillRowley:2012kx}.  
Recently, it has been shown \cite{Bandyopadhyay:2014tha} that  the fine-tuning can be reduced significantly when the TESSM is considered, since the extended Higgs sector can generate additional tree level contributions to the light Higgs mass so that sizable quantum corrections are no longer needed \cite{Delgado2013zfa,tessmh2,Bandyopadhyay:2014tha}.  For the experimentally viable parameter regions of the model it is also possible to obtain  an enhancement or a suppression in the Higgs decay to diphoton rate \cite{Bandyopadhyay:2014tha,rho,tessmh,Arina:2014xya}.  

The other important feature of the TESSM is that the three charged Higgs pairs in the model can have different collider signatures compared to the charged Higgses originated from the Higgs doublets. 
For the triplet dominated  charged Higgses, the production through the top decay or gg/gb fusions are no longer the most important channels at the LHC since triplet does not directly couple to the SM fermions. 
Though the charged Higgs sector of the TESSM has been partly studied in the literature \cite{BarradasGuevara:2004qi,chargedH}, all the possible production/decay channels have to be investigated to understand the possibility to observe a triplet-like charged Higgs at the LHC. 
In this spirit, this study is dedicated to examine the possible decay and the production channels for the charged Higgses where the impact of $h^{\pm}ZW^{\mp}$ vertex is not negligible. 
During the analysis, we take into account the possible charged Higgs production channels such as the pair production $h_i^{\pm}h_j^{\pm}$, the associated gauge boson production $h_i^{\pm}Z$, $h_i^{\pm}W^{\mp}$  and the associated neutral Higgs production $h_i^{\pm}h_j$, $h_i^{\pm}A_j$, along with the gg and gb fusions. 

Recent limits on the light charged Higgs at the LHC for 2HDM and MSSM have been obtained by considering  $pp\to t\bar{t}$ with $t\rightarrow h^{\pm}b$ production channel and charged Higgs decay  to $\tau \nu_{\tau}$ mode. The mass exclusion limits for the light charged Higgs have been attained by using $\mathcal{B}(t \to b h^\pm) \times \mathcal{B}(h^\pm \to \tau \nu)$ whereas for heavier charged Higgs bounds have been given for the effective production cross-section\cite{chbnd}.  Certainly for the models with non-standard decays of the light charged Higgs, the present mass bounds are less tight and many regions of light charged Higgs can be still allowed by the data \cite{chnsmssm}. In this article we select four benchmark points to study and probe the triplet like charged Higgs ($m_{h^\pm_i} \sim 200$ GeV) . We select our parameter points such that the triplet nature in the lightest charged Higgs allows  $ZW^\pm$ decay which can be as strong as  $tb$ and $\tau \nu_{\tau}$ modes when it is kinematically possible.  A Pythia-Fastjet level simulation is carried  out by taking care of dominant SM backgrounds. We analyse the signal in multi-lepton final states, i.e., with $3\ell$, $4\ell$, $5\ell$ along with $\tau$ and $b$-jets. The requirements of multi-lepton and jets reduce the background substantially and earliest discovery hint can come with $\gtrsim 72$ fb$^{-1}$
of integrated luminosity.  

The rest of the paper is organized as follows. In the next Section, we briefly review the model focusing especially on the Higgs sector where we give the effective formula describing the radiative corrections to the neutral Higgs masses. In Section \ref{chargedsection}, we investigate the features of the charged Higgses and especially focus on the  tree level $h_i^{\pm}ZW^{\pm}$ coupling. 
We construct three different scenarios to show the impact of $h_i^{\pm}\rightarrow ZW^{\pm}$  decay on the branching ratio of the lightest charged Higgs. Then we discuss possible channels that can contribute to the charged Higgs production at the LHC. In Section \ref{benchmarks}, we introduce four different benchmark points with corresponding charged Higgs production cross-sections. In Section \ref{collidersimulation} we perform the collider simulations and  discuss different final states. In Sections \ref{discussre} and \ref{conclusionsc}, we give our final remarks and conclusion. 
 
\section{The Model}\label{THM}

In TESSM, the field content of the MSSM is enlarged by introducing an $SU(2)$ complex Higgs triplet with zero hypercharge which can be represented as a 2x2 matrix 

 \begin{equation}
 {\bf T} = \begin{pmatrix}
       \sqrt{\frac{1}{2}}T^0 & T_2^+ \cr
      T_1^- & -\sqrt{\frac{1}{2}}T^0
       \end{pmatrix} .
 \end{equation}
Here $T^0$ is a complex neutral field, while  $T_1^-$ and $T_2^+$ are the charged Higgs fields. 
Note that $(T_1^-)^*\neq -T_2^+$.  The superpotential of  the Higgs sector of the model
is given by 
\begin{equation}
W=\lambda  \hat H_d \cdot \hat T  \hat H_u\, +\, \mu_D  \hat H_d \cdot  \hat H_u \, +\, \mu_T Tr(\hat T \hat T)+y_t \hat U \hat H_u\!\cdot\! \hat Q - y_b \hat D \hat H_d\!\cdot\! \hat Q- y_\tau \hat E \hat H_d\!\cdot\! \hat L\ ,
\label{superpot}
 \end{equation}
where $\mu_D$ is the usual mixing parameter of the two Higgs doublets and $\mu_T$ is the mass parameter of the triplet. The triplet field $T$  couples to the two Higgs doublets by a dimensionless 
coupling $\lambda$ and the triplet-SM fermion couplings are absent. 
We neglect other than the third generation Yukawa couplings.
Note that `` $\cdot$ ''  represents  contraction with antisymmetric $\epsilon_{ij}$ where $\epsilon_{12}=-1$. The soft SUSY breaking potential of the Higgs sector $V_S$ can be written by using the convention of the superpotential as
\be
V_S&= &m_{H_d}^2 |H_d|^2 + m_{H_u}^2 |H_u|^2 + m_{T}^2 {\rm Tr} (T^\dag T)
+[ B_D \mu_D H_d \cdot H_u + B_T \mu_T {\rm Tr} (T T)\\\nonumber
&&+A_{\lambda} \lambda H_d \cdot T H_u + y_t A_t \tilde{t}^*_R H_u\!\cdot\! \tilde{Q}_L- y_b A_b \tilde{b}^*_R H_d\!\cdot\! \tilde{Q}_L +{\rm h.c.} ] .\
\ee
Here $A_{j}$ ($j=\lambda,t,b$) are the soft trilinear parameters, $B_D$ and $B_T$ are the soft bilinear parameters, while $m_i$ ($i=H_d,H_u,T$) represent the soft SUSY breaking masses. 
For simplicity we assume that there is no CP violation in the Higgs sector and all the parameters as well as the vacuum expectation values of the neutral Higgs fields (vevs) are chosen to be real. 
When these neutral fields acquire non-zero vevs, denoted by
\be
\langle H_u^0 \rangle = \frac{v_u}{\sqrt{2}},\;\; \langle H_d^0 \rangle = \frac{v_d}{\sqrt{2}},\;\; \langle T^0 \rangle = \frac{v_T}{\sqrt{2}}, 
\ee
and $\tan{\beta}={v_u}/{v_d}$, the electroweak symmetry is spontaneously broken and all fermions and gauge bosons gain masses. The $W$ boson mass expression is altered by the triplet vev as 
$m_W^2=g_2^2(v^2+4v_T^2)/4$, where $v^2=v_u^2+v_d^2$; whereas the $Z$ boson mass expression remains unaffected. This non-zero triplet contribution to $W$ mass leads to a deviation in the tree-level $\rho$ parameter expression,
\bea
\rho=1+4v_T^2/v^2.
\eea
Thus the triplet vev is strongly constrained by the global fit on the $\rho$ 
parameter measurement \cite{pdg},
\bea
\rho=1.0004{\small{\begin{array}{l} + 0.0003\\-0.0004 \end{array} }}
\eea
which implies $v_T\leqslant5$ GeV. Such a stringent constraint on the triplet vev prevents   the triplet extension from generating $\mu_D$ term effectively as a solution for  $\mu$ problem of the MSSM. This is why we write the $\mu_D$ term and the triplet-doublet interaction term separately in the superpotential. However, small $v_T$ can still generate deviations in the Higgs sector as we show in the following Sections. 
In the numerical analysis of this paper we use a fixed value $v_T=3\sqrt{2}$ GeV for the triplet vev. 
 
\subsection{The Higgs Sector of the TESSM}

After the electroweak symmetry breaking the physical particle spectrum of the TESSM Higgs sector comprises three CP-even ($h_{1,2,3}$), two CP-odd ($A_{1,2}$) and three charged Higgs bosons ($h^{\pm}_{1,2,3}$). In our notation, $h_1$ corresponds to the lightest Higgs boson of the model whereas the others are generally much heavier. It was shown in Ref.~\cite{rho,Bandyopadhyay:2014tha} that it is possible to obtain the lightest Higgs boson with a mass up to $125$ GeV even at tree level when $\lambda$ is so large that the model is in the non-perturbative regime at the high scales. 
As long as we demand perturbativity of the model, for a $125$ GeV Higgs one needs radiative corrections to the neutral Higgs sector which, however, are not necessarily as large as those in the MSSM \cite{tessmh2,tessmh,Delgado2013zfa,chargedH,Bandyopadhyay:2014tha}. 
The additional contributions to Higgs mass therefore decreases the required limits on the third generation squarks significantly \cite{tessmh2} and the corresponding fine-tuning can be greatly reduced as compared to values attainable in MSSM \cite{Bandyopadhyay:2014tha}. The one-loop radiative corrections to the Higgs potential can be calculated using the effective potential approach \cite{coleman-weinberg}

\bea
\Delta V=\frac{1}{64\pi^2}{\rm Str}\left[\mathcal{M}^4\left(\ln\frac{\mathcal{M}^2}{\Lambda^2}-\frac{3}{2}\right)\right].
\eea
Here $\Lambda$ is the renormalization scale and $\mathcal{M}$ represents the field dependent mass matrices  that can be found in Ref. \cite{Bandyopadhyay:2014tha}.  
In the current study, we consider the radiative corrections coming both from the colored sector and the electroweak sector since it is known that for TESSM the neutralino and chargino corrections can have crucial impact on the lightest Higgs mass \cite{rho,Bandyopadhyay:2014tha,tessmh2}. For the numerical calculations, we use an effective formula for the 1-loop contributions to the CP even scalar mass matrix  \cite{rho,Bandyopadhyay:2014tha} 

\begin{align}
(\Delta\mathcal{M}^2_{h^0})_{ij}
&=\left.\frac{\partial^2\Delta V(a)}{\partial a_i\partial a_j}\right|_{\rm{vev}}
-\frac{\delta_{ij}}{\langle a_i\rangle}\left.\frac{\partial \Delta V(a)}{\partial a_i}\right|_{\rm{vev}}
\label{1Lmha}\\
&=\sum\limits_{k}\frac{1}{32\pi^2}
\frac{\partial m^2_k}{\partial a_i}
\frac{\partial m^2_k}{\partial a_j}
\left.\ln\frac{m_k^2}{\Lambda^2}\right|_{\rm{vev}}
+\sum\limits_{k}\frac{1}{32\pi^2}
m^2_k\frac{\partial^2 m^2_k}{\partial a_i\partial a_j}
\left.\left(\ln\frac{m_k^2}{\Lambda^2}-1\right)\right|_{\rm{vev}}
\nonumber\\
&\quad-\sum\limits_{k}\frac{1}{32\pi^2}m^2_k
\frac{\delta_{ij}}{\langle a_i\rangle}
\frac{\partial m^2_k}{\partial a_i}
\left.\left(\ln\frac{m_k^2}{\Lambda^2}-1\right)\right|_{\rm{vev}}\ ,\quad i,j=u,d,T\  .
\label{1Lmh}
\end{align}
Here $m^2_k$ is the set of eigenvalues of the field dependent mass matrices given in Ref. \cite{Bandyopadhyay:2014tha} where the real components ($a_j$) of the neutral Higgs fields are defined as $H_u^0=1/\sqrt{2}\left(a_u+i b_u\right)$, $H_d^0=1/\sqrt{2}\left(a_d+i b_d\right)$ and $T^0=1/\sqrt{2}\left(a_T+i b_T\right)$. For simplicity we drop the supertrace expressions in Eq. (\ref{1Lmh}) but for each particle, the supertrace coefficient should be taken into account.  

In the rest of the paper we particularly focus on the charged Higgs sector of the model.
The charged Higgs in TESSM has been partly investigated in the literature earlier \cite{BarradasGuevara:2004qi,chargedH}. 
Further studies are needed regarding the mass spectrum and the LHC signatures of the charged Higgses to pin down the structure and symmetries of the underlying Higgs sector. 
In this sprit, we perform a scan for the viable points containing one $\sim 125$ GeV neutral Higgs in the spectrum and discuss the features of the charged Higgs sector of TESSM in the next section.  


\section{The Charged Higgs Phenomenology}
\label{chargedsection}

\begin{figure}[thb]
\mbox{\subfigure[]{\includegraphics[width=0.48\linewidth]{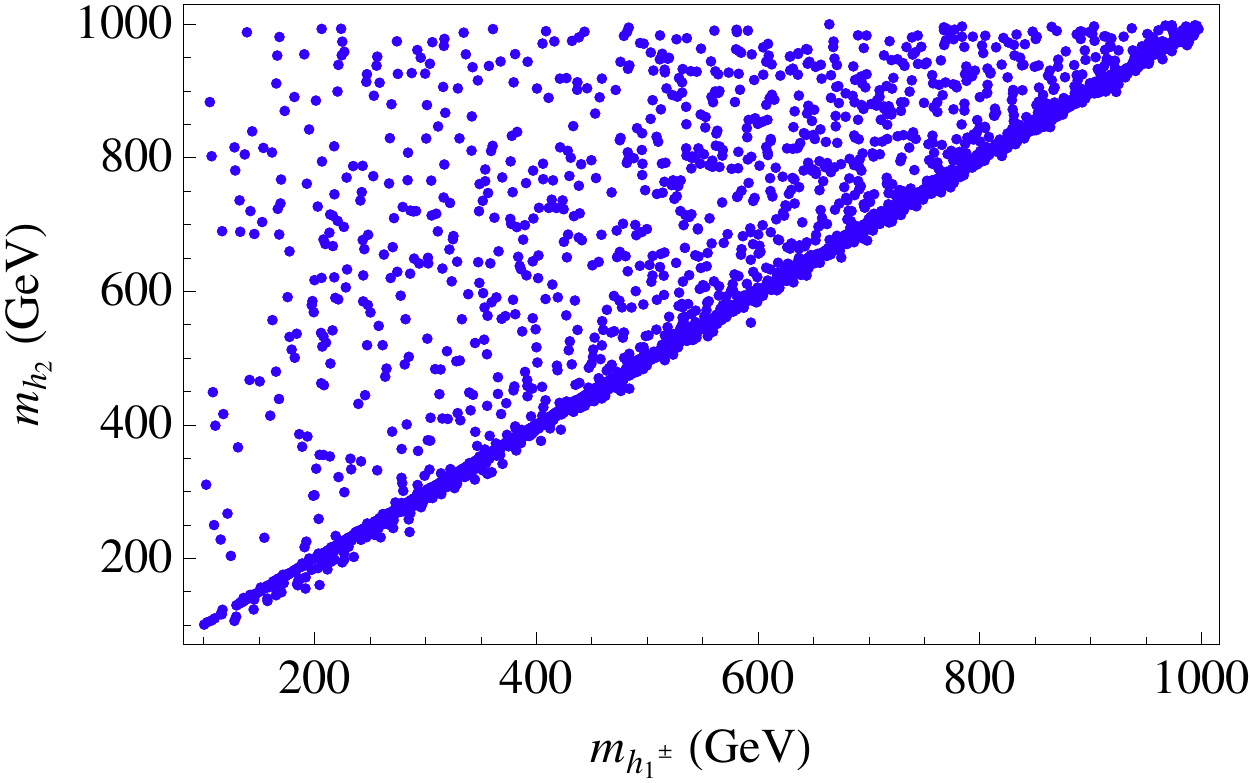}}
\subfigure[]{\includegraphics[width=0.48\linewidth]{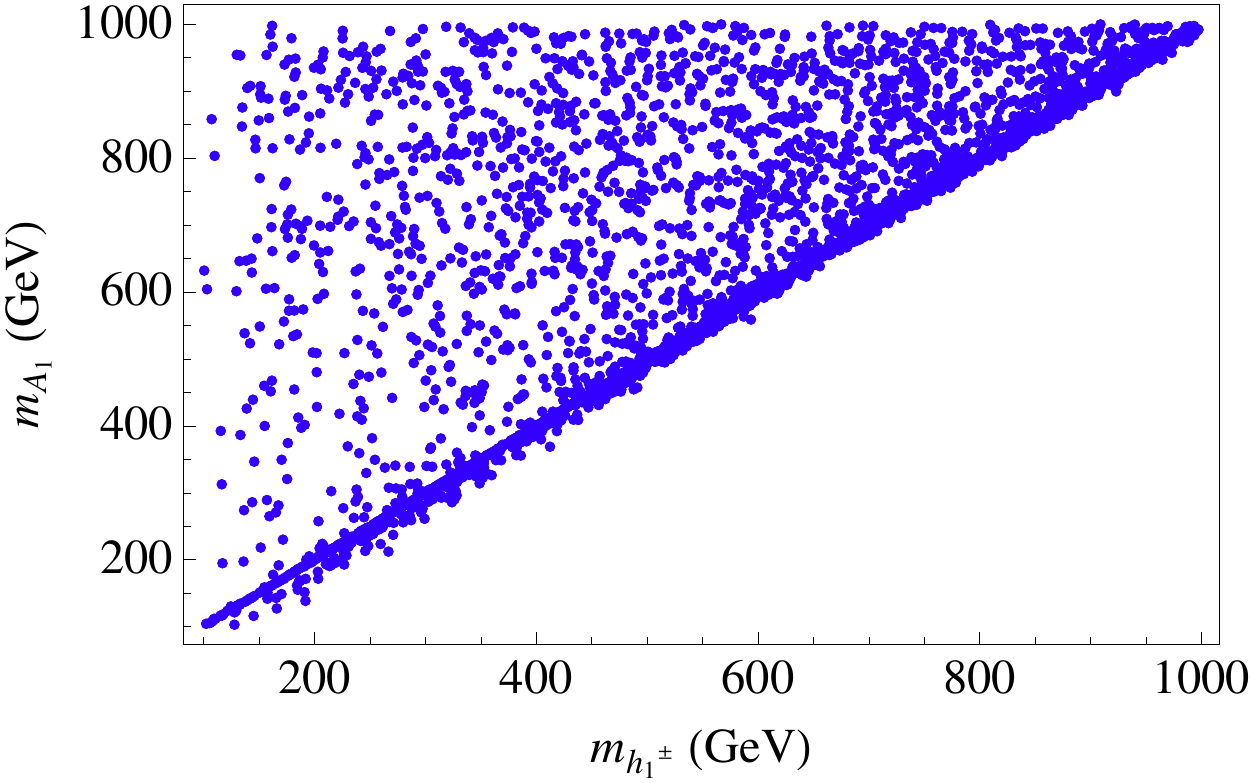}}}
\caption{The mass hierarchy between the lightest charged Higgs boson and  (a) the heavier CP even Higgs $h_2$ (b) the CP odd Higgs boson $A_1$.}\label{higgsspect}
\end{figure}

The charged Higgs boson is a clear indication of a non-standard Higgs sector since SM predicts only one neutral scalar boson in the particle spectrum. The existence of at least one charged Higgs is predicted by many models with different multiplet structure and probing the properties of charged Higgs can differentiate between the models. For this purpose, we first investigate the mass hierarchy among the Higgs bosons of TESSM to distinguish the model from the ones containing only Higgs doublets in their field contents. To present the properties of the TESSM Higgs sector, we perform a scan for which the parameter ranges are defined as:
\begin{eqnarray}\label{pscan}
&&1\leq {\tan\beta }\leq 30,\  \left|\lambda\right|\leq1,\,0\leq \left|\mu _D,\mu _T\right|\leq 2 \,\text{TeV},\ 100 \,\text{GeV}\leq \left|M_1,M_2\right|\leq 1  \,\text{TeV},\nonumber\\ 
&& 0\leq \left| A_t,A_b,A_{\lambda},B_D,B_T\right|\leq 2 \,\text{TeV},\ 500 \,\text{GeV}\leq m_Q,m_{\tilde{t}},m_{\tilde{b}}\leq 2\,\text{TeV} \end{eqnarray}
where $m_Q,m_{\tilde{t}},m_{\tilde{b}}$ are the left- and right-handed squark soft masses and $M_i$ (i=1,2) are the soft gaugino masses. In this parameter scan, we collect the random parameter points that respect the following constraints
\begin{eqnarray}
124\leq m_{h_1^0}\leq 127\, {\rm GeV}\ ;\ m_{A_{1,2}},\ m_{\chi^0_{1,2,3,4,5}}&\geq & 65\,{\rm GeV}\ ;\nonumber\\
 \,m_{\chi^\pm_{1,2,3}}\geq 104\,{\rm GeV} \ ; \ m_{\tilde{t}_{1,2}},m_{\tilde{b}_{1,2}}&> & 600\,{\rm GeV}\ .
\end{eqnarray}
Here our bound for neutralinos is stronger than the experimental one in order to avoid the constraints from the invisible decays of the lightest Higgs boson.
Such decays are relevant for the dark matter studies \cite{Arina:2014xya}. In Fig. \ref{higgsspect}a (b) we display the mass hierarchy between the lightest charged Higgs boson and the heavier CP even Higgs boson $h_2$ (CP odd $A_1$) for the viable points respecting the aforementioned constraints. 
Unlike in the case of MSSM, often the lightest charged Higgs is non-degenerate in mass with CP odd and heavy CP even Higgs bosons. Such non-degeneracy has a great importance since on-shell decays like $h_2 (A_1)\rightarrow h_1^{\pm}W^{\pm}$ are now kinematically possible and can give significant contribution to the charged Higgs production cross-section.  We also observe that for the obtained parameter points, the heaviest CP odd ($A_2$) and CP even ($h_3$) Higgs bosons have mass $\gtrsim 500$ GeV.
Their negligible contributions to the total production cross-section are not considered during the collider simulations.

In addition to non-degeneracy, the charged Higgs couplings to SM fermions and gauge bosons must be revisited to understand the triplet impact. It is known that in two Higgs doublet models (with or without supersymmetry), $h^{\pm}ZW^{\mp}$ coupling can be induced only at loop order since the custodial $SU(2)_c$ symmetry prohibits the tree level interaction. However, for $Y=0,\pm 2$ triplet extended models, the custodial symmetry is broken by the triplet vev that allows non-zero $h_{i}^{\pm}ZW^{\mp}$ coupling at tree level. This coupling has a great importance since its verification could indicate the evidence of higher multiplet structure in the electroweak sector. For the $Y=0$ triplet extended model this coupling can be written as
\bea
g_{h_{i}^\pm W^\mp Z}=-\frac{1}{2} i g_2 \left[g_1 \sin\theta_w (v_u R_{(i+1)1}-v_d R_{(i+1)2})+\sqrt{2} g_2 v_T \cos\theta_w (R_{(i+1)3}+R_{(i+1)4})\right],\label{h+zw-coup}
\eea 
where $g_1$ and $g_2$ are the electroweak gauge couplings and $\theta_w$ is the Weinberg angle. ${R}_{ij}$ are the rotation matrix entries of charged Higgs sector and the physical charged Higgses  are related with the charged fields as 

\begin{eqnarray}
\left( \begin{array}{c}
G^+\\
h_1^{+}\\
h_2^+\\
h_3^+\\
\end{array}\right)
\label{rotmatrix}=\left( \begin{array}{cccc}
R_{11}&R_{12}&R_{13}&R_{14}\\
R_{21}&R_{22}&R_{23}&R_{24}\\
R_{31}&R_{32}&R_{33}&R_{34}\\
R_{41}&R_{42}&R_{43}&R_{44}\\\end{array}\right)
\left( \begin{array}{c}
H_u^+\\
H_d^{-*}\\
T_2^+\\
T_1^{-*}
\end{array}\right).
\label{rotmatrix}
\end{eqnarray}
During the investigation of the $h_{i}^\pm W^\mp Z$ coupling strength, the numerically calculated $4\times 4$ rotation matrix entries require special attention. When one of the $R_{ij}$ entries is set to a value, the other entries must be chosen carefully so as to respect the electromagnetic gauge symmetry that ensures $h_{i}^{\pm}W^{\pm}\gamma$ coupling to be zero.

\begin{figure}[thb]
\centering
\includegraphics[width=0.5\linewidth]{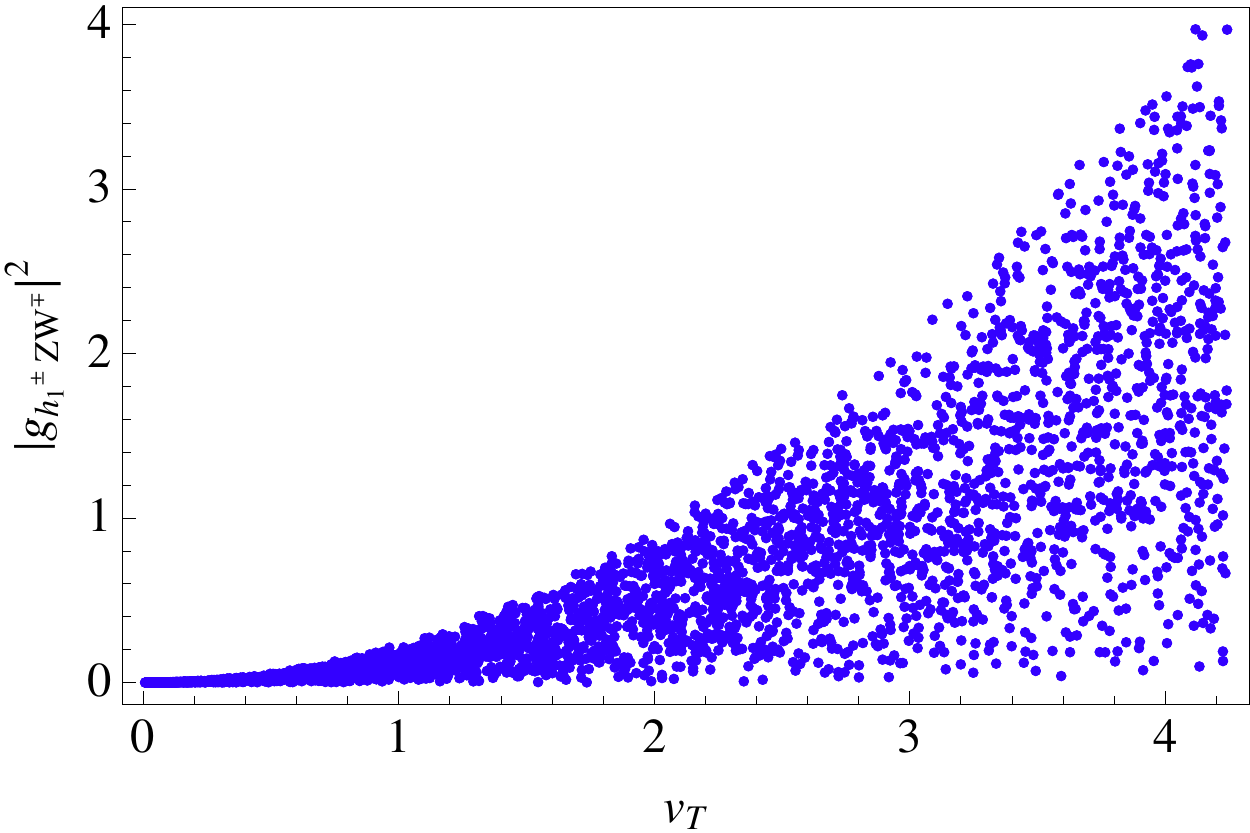}
\caption{$|g_{Zh_1^{\pm}W^{\mp}}|^2$ as a function of $v_T$: For $v_T=0$ the coupling vanishes due to the $U(1)_{em}$ gauge invariance.}\label{ZHWvt}
\end{figure}
In order to see the strength of the coupling we perform another scan where we consider $1\leq\tan\beta\leq30$ and display the coupling strength $|g_{Zh_1^{\pm}W^{\mp}}|^2$ as a function of $v_T$ in Fig. \ref{ZHWvt}. 
When $v_T$ is non-zero the first term\footnote{The first term was not considered in Ref. \cite{BarradasGuevara:2004qi}.} contributes along with the second term and according to the doublet-triplet content in the charged Higgs, the coupling can be suppressed or enhanced. However, when $v_T$ is zero the coupling vanishes since expressions in the parenthesis in the first part of the coupling must cancel each other not to lead to the tree level $h_i^{\pm}W^{\mp}\gamma$ coupling that breaks the electromagnetic gauge invariance. In other words, $h_{i}^{\pm}ZW^{\pm}$ can survive only for non-zero triplet vacuum $v_T$. Similar situation arises when the triplet is completely decoupled from the doublet sector. For the doublet like lightest charged Higgs the second term in the coupling again drops and the two terms in the first part cancels exactly like in the case of MSSM. To understand the impact of this tree level coupling on the charged Higgs decays, the next section is devoted to the study on the light charged Higgs branching ratios where we construct different scenarios to reveal the triplet effect on the decay channels.

\subsection{Charged Higgs Decay Channels}\label{chdcy}

We now discuss the decay of the lightest charged Higgs boson in the low mass region $150<m_{h_1^{\pm}}\lesssim 200$ GeV. To show the impact of the triplet nature, we construct three scenarios where charged Higgs has the different percentage of triplet component. During the numerical analysis, we set $R_{23}=0.3$ in the $h_{1}^{\pm}ZW^{\mp}$ coupling and we determine the other $R_{ij}$ parameters ensuring that the $h_1^{\pm}W^{\mp}\gamma$ coupling is zero at tree level. 
For the numerical analysis we select the following scenarios:

\begin{itemize}
\item {\it {Scenario I} (Sc I):} The rest of the $R_{ij}$ elements in the $h_{1}^{\pm}ZW^{\mp}$ coupling are arranged in such a way that the triplet component in the lightest charged Higgs is $90\%$.

\item {\it {Scenario II} (Sc II):} The rest of the $R_{ij}$ elements are chosen for the lightest charged Higgs to comprise $1\%$ doublet and $99\%$ triplet.

\item {\it {Scenario III} (Sc III):} The charged Higgs is composed of $0.1\%$ doublet and $99.9\%$ triplet.\end{itemize}

\begin{figure}[thb]
\mbox{
\subfigure[]{\includegraphics[width=0.5\linewidth]{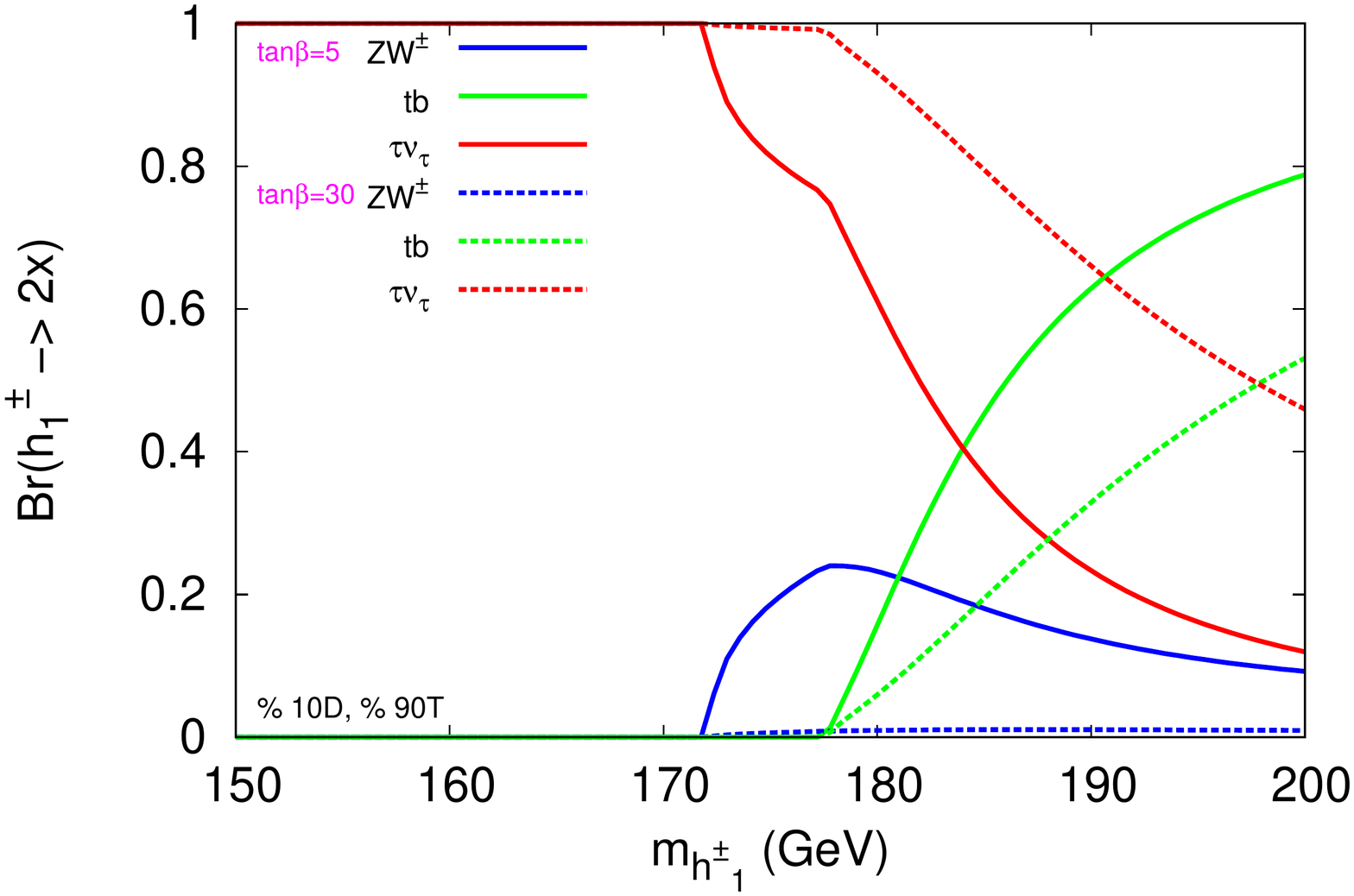}}
\subfigure[]{\includegraphics[width=0.5\linewidth]{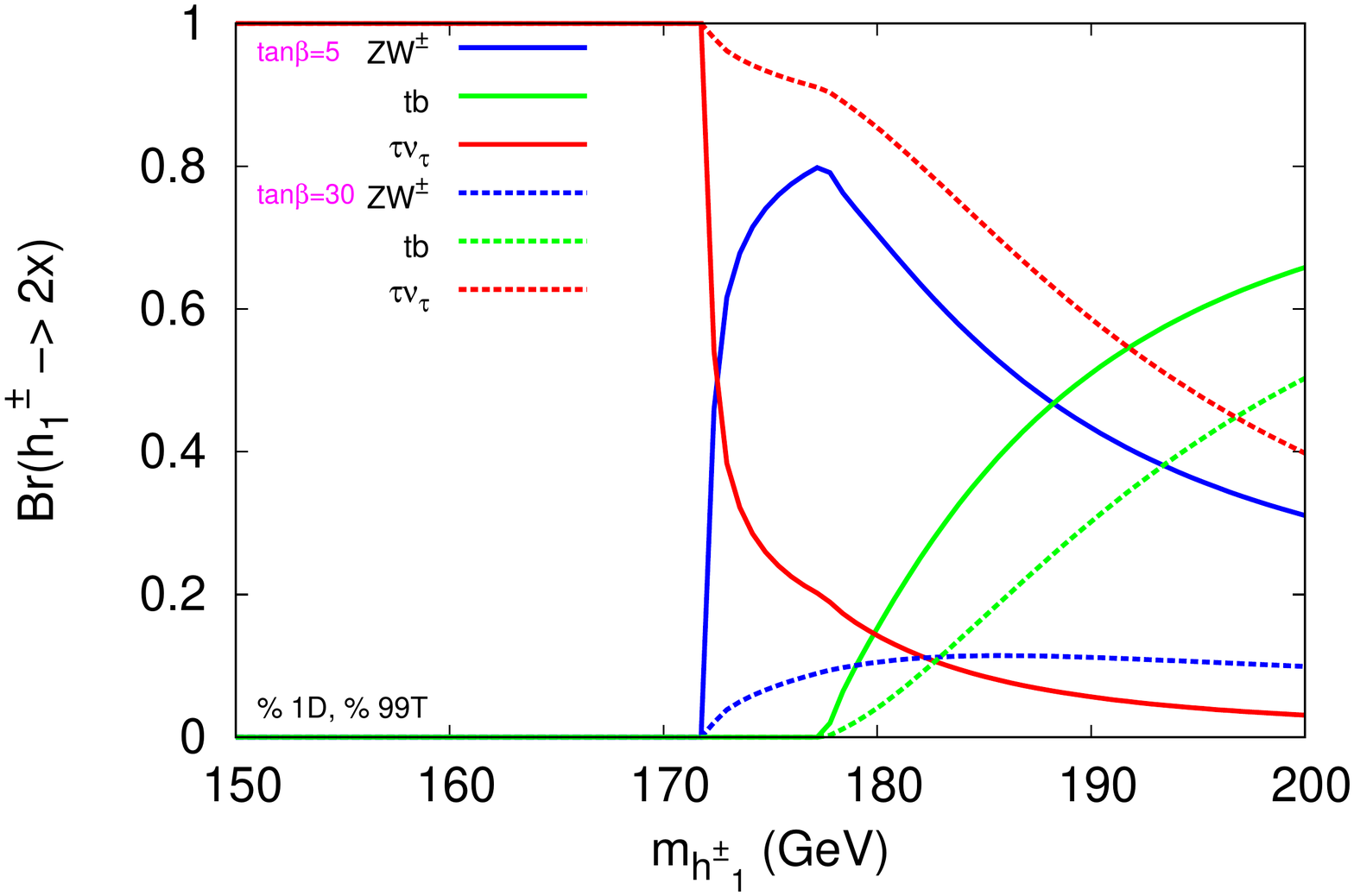}}}
\centering
\mbox{
\subfigure[]{\includegraphics[width=0.5\linewidth]{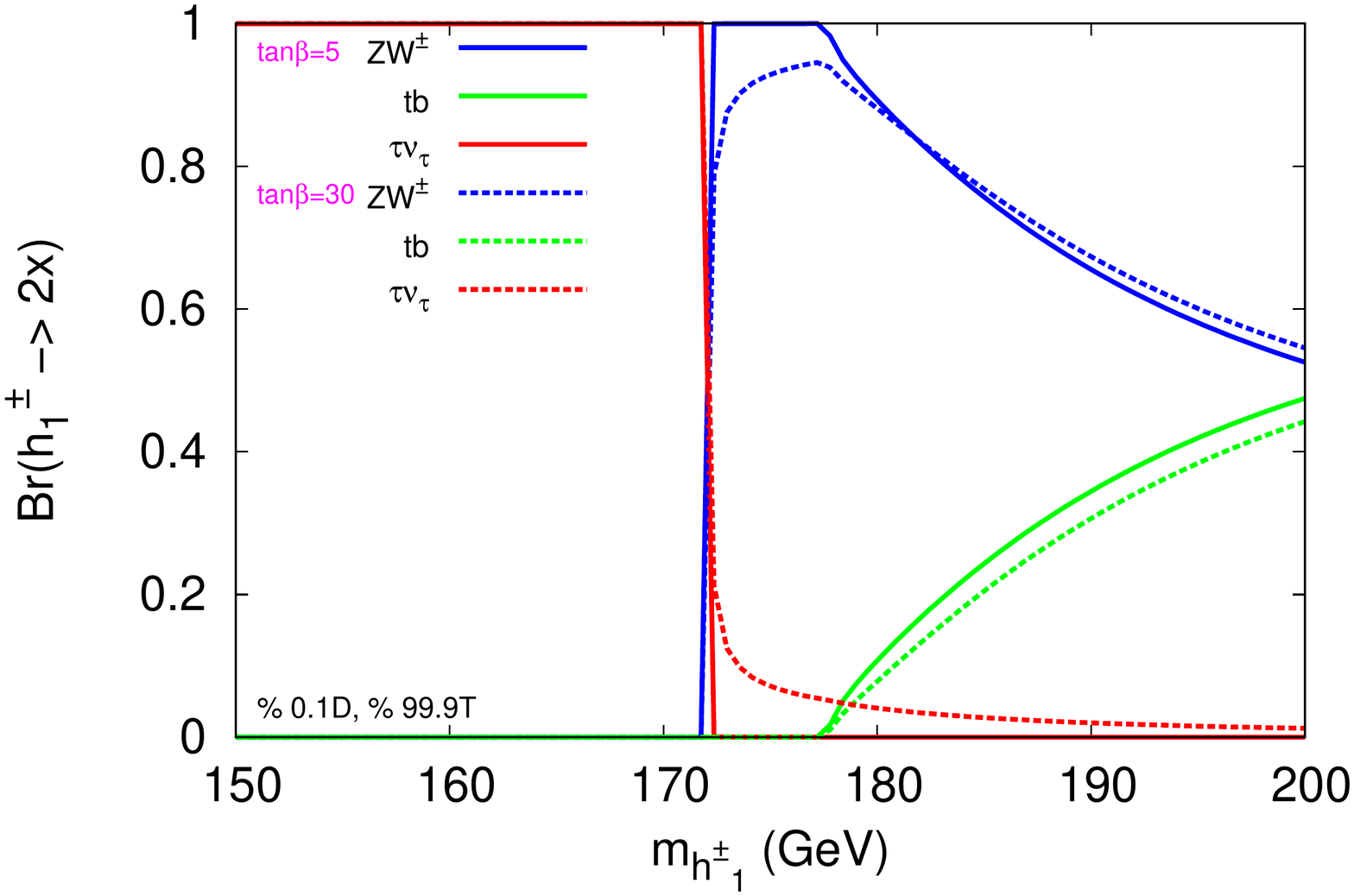}}}
\caption{The branching ratios of the lightest charged Higgs for (a) Scenario I (b) Scenario II and (c) Scenario III. }\label{Brchsc1-4}
\end{figure}

To determine the branching ratios of the lightest charged Higgs in TESSM we employ \texttt{SARAH} \cite{sarah1,sarah2} program to generate the model files for \texttt{Calchep} \cite{calchep} and we display the results in Fig.\ref{Brchsc1-4}. 
For this analysis we choose $\tan\beta=5$ and $30$ to investigate the low and high $\tan\beta$ behaviors for all scenarios. 
When the doublet component of the lightest charged Higgs is $10\%$ (Scenario I) for $\tan\beta=5$, $\mathcal{B}r(h_1^{\pm}\rightarrow Z W^{\pm})$ can reach up to $25 \%$ for $m_h^{\pm}\sim 175$ GeV (See Fig.~\ref{Brchsc1-4}a).
However, as soon as $tb$ channel is kinematically possible the branching ratio to $ Z W^{\pm}$ drops to $10 \%$. 
For $\tan\beta=30$ for $m_{h_1^{\pm}}\leq m_t+m_b$ region the dominant decay channel is $\tau \nu_{\tau}$ and $ZW^{\pm}$ channel is very suppressed. For a heavier charged Higgs mass, $tb$ channel becomes as important as $\tau\nu_{\tau}$ whereas the $ZW^{\pm}$ channel has an insignificant branching. 

This behavior changes completely in the case of Sc II. 
When the triplet component increases in the lightest charged Higgs, its coupling to fermions diminishes since triplet does not couple to the SM fermions directly and thus the fermionic branching ratios drop significantly. 
For $\tan\beta =5$, the branching ratio to $ZW^{\pm}$ increases dramatically and it can be as high as $80 \%$ (See Fig.~\ref{Brchsc1-4}b). 
When $tb$ channel is open the $ZW^{\pm}$ branching ratios drops to $\sim 30 \%$ but it still much larger than $\tau\nu_{\tau}$ branching ratios. 

In Sc III, the lightest charged Higgs is almost triplet and $ZW^{\pm}$ channel becomes the most dominant decay channel even after $tb$ decay is kinematically possible. The branching can be as high as 100$\%$ for $\tan\beta=5$ and it drops to $\sim 50 \%$ for $m_{h_1^{\pm}}=200$ GeV (Fig.~\ref{Brchsc1-4}c). For $\tan\beta=30$ the situation is similar except that the branching ratio can reach up to $\sim 90 \%$ for the mass region $170\leq m_{h_1^{\pm}} \leq 180$ GeV.

Next we focus on triplet effect on the charged Higgs production channels at LHC and we discriminate the dominant charged Higgs production channels for TESSM from the ones in the isospin doublet models.

\subsection{Charged Higgs Production Channels at the LHC}\label{chprod}

\begin{figure}
\centering
\mbox{\subfigure[]{\includegraphics[width=2.1in,height=2.1in]{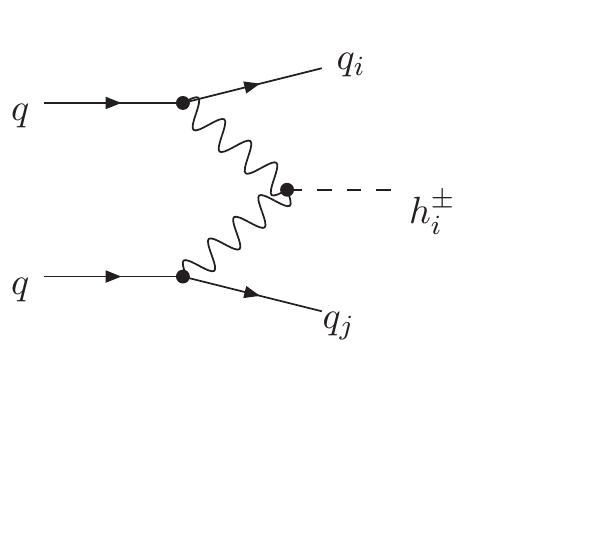}}
\subfigure[]{\includegraphics[width=2.1in,height=2.1in]{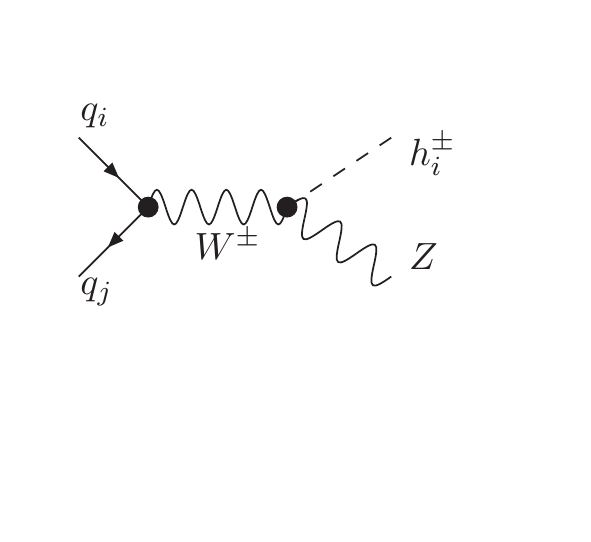}}
\subfigure[]{\includegraphics[width=2.1in,height=2.1in]{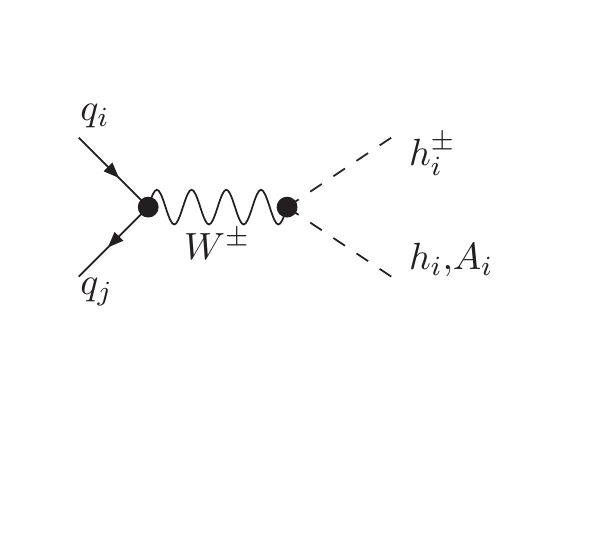}}}
\mbox{
\subfigure[]{\includegraphics[width=2.1in,height=2.1in]{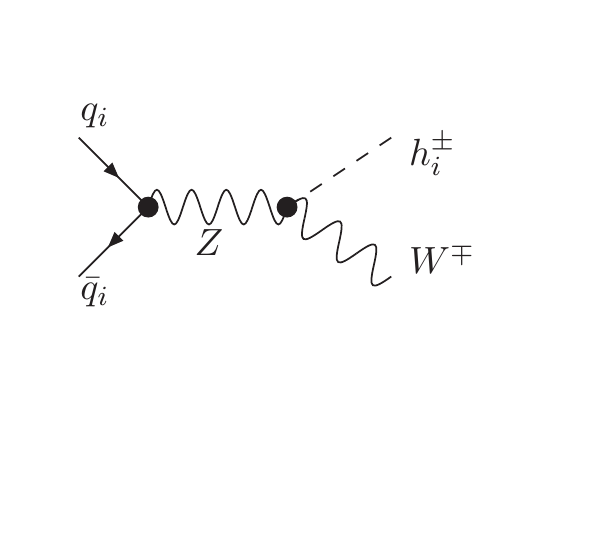}}
\subfigure[]{\includegraphics[width=2.1in,height=2.1in]{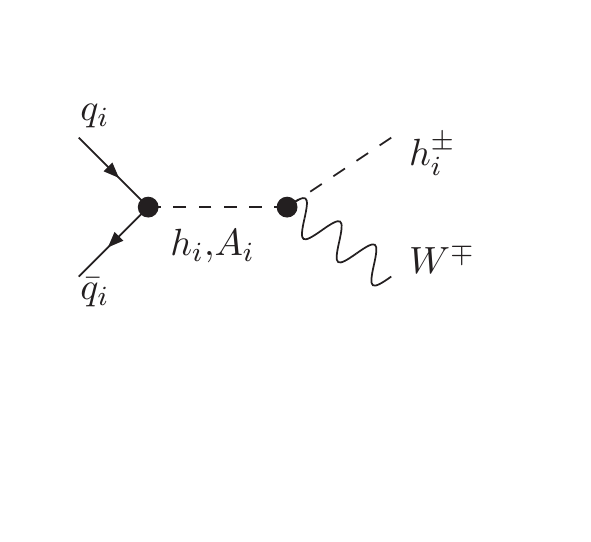}}}
\caption{ The main production channels of triplet like charged Higgses at LHC in the TESSM.}\label{h+main_pro}
\end{figure}

The production channels for the charged Higgses depend strongly on the model under consideration. 
When the model has only Higgs doublets ({\it e.g.} 2HDM, MSSM), the main production channel of the charged Higgs with mass $m_{h_i^{\pm}}\leq m_t-m_b$ is $pp\to t\bar{t}$ and the subsequent top quark decay to $b$ and the charged Higgs boson. 
For $m_{h_i^{\pm}}\geq m_t+m_b$ the dominant channels are the associated production with a top quark in $gb$ and $gg$ fusions \cite{Gunion:1986pe,Barger:1993th,DiazCruz:1992gg,Berger:2003sm}. 
The $q\bar{q}$ annihilation also contributes non-negligibly to the latter case. However $gg\rightarrow tb h_i^{\pm}$ and $gb (g\bar{b})\rightarrow t h_i^{-} (\bar{t}h_i^{+})$ channels are excepted to be suppressed in TESSM since the triplet nature  abolishes the strength of charged Higgs-SM fermion couplings. These channels could be still non negligible as long as the charged Higgs is a mixture of doublet and triplet fields. For $q\bar{q}$ annihilation cross-section similar suppression occurs in the contributions containing direct quark-charged Higgs coupling. However,  the additional contributions to the production cross-section in question may arise due to the tree level $h_i^{\pm}ZW^{\mp}$ interaction and the $h_i^{\pm}$ production becomes possible through the vector boson fusion (Fig. \ref{h+main_pro}a).

\begin{figure}
\centering
\mbox{
\subfigure[]{\includegraphics[width=2in,height=2.3in]{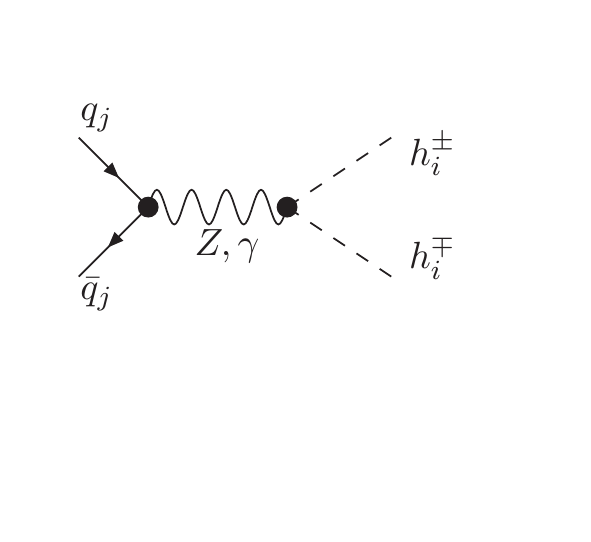}}
\subfigure[]{\includegraphics[width=2in,height=2.3in]{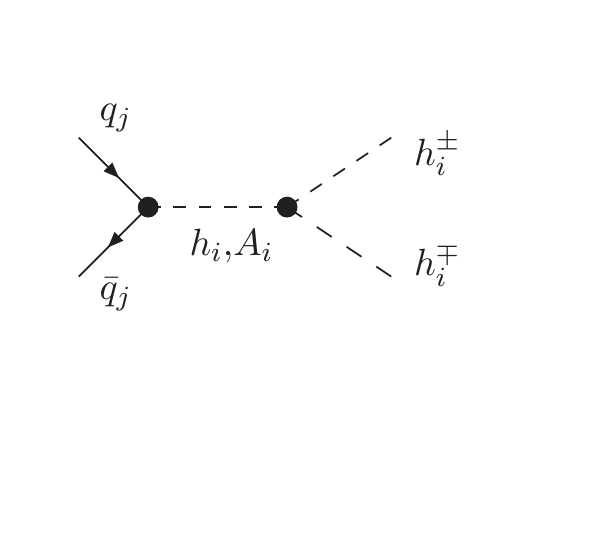}}
\subfigure[]{\includegraphics[width=2in,height=2.3in]{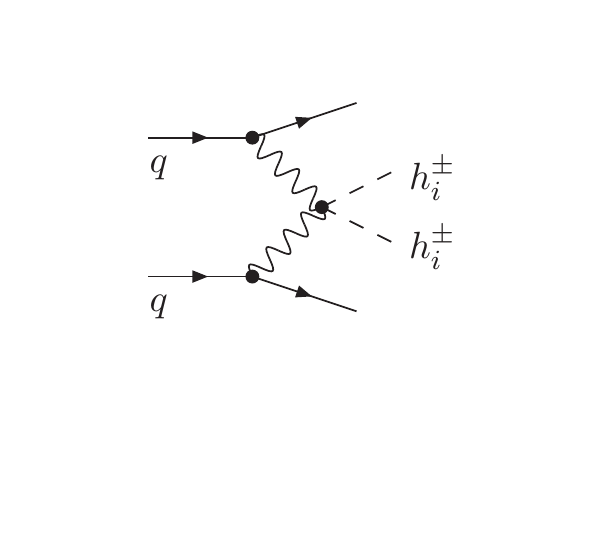}}}
\caption{ Some example diagrams for the charged Higgs pair production at LHC for TESSM.}\label{h+h-production}
\end{figure}

The other interesting alternatives to produce $h_i^{\pm}$ bosons are the pair production and the associated production mechanisms with massive gauge bosons as well as the neutral Higgs bosons. In Fig. \ref{h+main_pro}b we display the s-channel $h_i^{\pm}$ production in association with $Z$ through $q\bar{q}$ annihilation. This channel is only possible thanks to tree level $h_i^{\pm}ZW^{\mp}$ vertex which can also enhance the production of the charged Higgs in association with $W^{\pm}$ boson (Fig. \ref{h+main_pro}d) in addition to the production through neutral Higgs propagations as shown in Fig. \ref{h+main_pro}e. We also take into account the associated production with neutral bosons (Fig. \ref{h+main_pro}c) since the TESSM Higgs sector contains more neutral Higgses than 2HDM or MSSM, and this affects the production cross-section in question. 
In Fig. \ref{h+h-production} we show some example diagrams for the charged Higgs pair production mechanisms that can give significant contributions to the signals that we investigate in collider simulation section. 
The triplet nature of the neutral Higgses can also play an important role since the processes in Fig. \ref{h+main_pro}e and Fig. \ref{h+h-production}b are expected to be suppressed for triplet like neutral Higgses. 
The analysis for determining the dominant production channel of the charged Higgs must be carried out carefully by investigating the nature of all Higgs bosons. 

In the following section, we consider four benchmark points with the triplet percentages in the neutral and charged Higgses and we determine the dominant production mechanism of the lightest charged Higgs for each benchmark point and study the possible multi-lepton signatures at the LHC.  

\section{Benchmark points and final states}
\label{benchmarks}
For the collider analysis we mainly consider phenomenology of the light charged Higgs with mass $m_{H^\pm}\sim 200$ GeV. 
The light charged Higgs scenario is very interesting for distinguishing the nature of the Higgs sector.  
The reason behind is  that in the models containing two doublets and/or singlets such a light charged Higgs boson mainly decays to either $tb$ or $\tau \nu$ whereas in models with SM-fermiophobic triplet such as TESSM these decays are expected to be suppressed. As discussed in the previous sections, for these models  $ZW^\pm$ decay channel can become as significant as the fermonic ones. To show the impact of the triplet nature on the decay and production channels of the charged Higgs here we consider four different benchmark points (BP's) with various doublet-triplet mixing in the Higgs sector.  During the analysis we focus on the low $\tan\beta$ region ($\tan\beta \leq 10$) which is favored by small fine-tuning \cite{Bandyopadhyay:2014tha}. In this region the tree level triplet contribution to the Higgs mass is maximized with large values of $\lambda$ \cite{Bandyopadhyay:2014tha,rho} so that as sizable loop corrections as in the case of MSSM are not needed. 
\begin{table}
\begin{center}
\renewcommand{\arraystretch}{1.4}
\begin{tabular}{||c|c|c|c|c|c|c|c||}
\hline\hline
Benchmark&$\tan\beta$&$m_{h_2}$&$m_{A_1}$& $m_{h_1^{\pm}}$& Doublet& Doublet &Triplet\\
Points & &(GeV)&(GeV)&(GeV)& $\%$ in $h_2$ &$\%$ in $A_1$ &$\%$ in $h_1^{\pm}$ \\ \hline\hline
BP1 &  8.63 & 182.898 & 610.91 & 182.942 & 1.34 & 99.967 & 98.88\\
\hline
BP2 & 4.89 & 216.94 & 451.453 & 216.41 & 0.2 & $<10^{-5}$ & 99.88 \\
\hline
BP3 & 6.32 & 441.507 & 198.438 & 197.854 & 0.12 & $<10^{-5}$ & 99.99 \\
\hline
BP4 & 7.23 & 362.843 & 184.706 & 183.637 & 0.78 & 0.006 & 99.98 \\
\hline
\hline
\end{tabular}
\caption{Benchmark points for a collider study consistent with the $\sim 125$ GeV Higgs mass where the $h_2$, $A_1$ and $h_1^{\pm}$ masses are calculated at tree level.}\label{bps}
\end{center}
\end{table}

Table~\ref{bps} shows the benchmark points chosen for our collider study in the next section. 
For our purposes, we consider the benchmark points that contain the lightest neutral Higgs with a mass calculated at one loop level satisfying the range $124\leq m_h \leq 127$ GeV. It is checked that all the benchmark points satisfy the experimental results for the Higgs decay rate to ZZ  and WW within $2 \sigma$ \cite{Chatrchyan:2013mxa,atlas1}. We also check that for all the benchmark points both heavy neutral Higgs bosons ($h_2$ and $A_1$)
respect the CMS \cite{hhcms} and ATLAS \cite{hhatlas} cross-section constraints at $2\sigma$ level. 
The lightest charged Higgs in all the four benchmark points has a substantial triplet component ($\gsim 99\%$) that enlarges the branching ratio of the $ZW^{\pm}$ channel as seen in the section \ref{chdcy}. 
We can see that either the lightest pseudo scalar $A_1$ (in BP3 and BP4) or the second lightest scalar $h_2$ (in BP1 and BP2) masses are nearly degenerate with the charged Higgs boson mass. The other heavier neutral Higgs bosons  $h_2$ (in BP3 and BP4) and $A_1$ (in BP1 and BP2) can decay to the lightest charged Higgs boson ($h^+_1$), unlike the cases of 2HDM and MSSM, where the heavier neutral Higgses ($A, H$) are almost degenerate with the charged Higgs boson in the spectrum.
Thus for the studied benchmark points the decay modes are drastically different than in MSSM case with heavy Higgs bosons.

\begin{table}
\begin{center}
\renewcommand{\arraystretch}{1.4}
\begin{tabular}{||c|c|c|c|c|c|c|c|c||}
\hline\hline
$h_2$ Decay&\multicolumn{8}{|c|}{Branching fractions(\%)}\\
\hline
Modes & $W^\pm h_1^\mp$ & $W^\pm W^\mp$ & $A_1$Z &\; $ZZ$ \; & \;$h_1h_1$\; & \;$b\bar{b}$ \;& \;$t\bar{t}$ \;&\;$\tau \bar{\tau}$ \\
\hline\hline
BP1 & - & 1.72 & - & 3.36 & - & 90.2 & - & 4.70 \\
\hline
BP2 & - & 55.0 & - & 43.6 & - & 1.33 & - & 6.93$\times10^{-2}$  \\
\hline
BP3 & 69.2 & 12.1 & 2.1$\times 10^{-9}$ & 6.71 & 8.15 & 0.0121 & 3.81 & 6.24$\times 10^{-4}$  \\
\hline
BP4 & 80.8 & 6.74 & 7.96$\times 10^{-3}$ & 3.97 & 5.01 & 3.08 & 0.201 & 0.160 \\
\hline
\end{tabular}
\caption{Decay branching fractions of $h_2$ for the benchmark points where the $h_2$ mass is calculated at tree level.
The kinematically forbidden decays are marked with dashes.}\label{H1dcy2}
\end{center}
\end{table}

Table~\ref{H1dcy2} presents the decay branching fraction for the second lightest CP-even scalar, $h_2$, where dash lines show the kinematically forbidden decays.  
We can see that in the case of BP1, $W^{\mp}h_1^{\pm}$ channel is not open and $h_2$ mainly decays to  $b\bar{b}$ due to its 1.34\% doublet component. For BP2,  the $b\bar{b}$ channel is very suppressed since $h_2$ is mostly triplet so it decays mainly to gauge boson pairs ($W^\pm W^\mp$ and $ZZ$). For BP3 and BP4, $h^\pm_1 W^\mp$ channel is kinematically open and it becomes the dominant decay channel due to the triplet nature of $h_2$. 

\begin{table}
\begin{center}
\renewcommand{\arraystretch}{1.4}
\begin{tabular}{||c|c|c|c|c|c|c|c|c||}
\hline\hline
$A_1$ Decay&\multicolumn{8}{|c|}{Branching fractions (\%)}\\
\hline
Modes & $W^\pm h_1^\mp$ & $\chi^\pm_1 \chi^\mp_1$ &\; $h_1$Z \;&\; $h_2$Z&\; $\chi_1^0 \chi_1^0$ \;  & \;$b\bar{b}$ \;& \;$t\bar{t}$ \;&\;$\tau \bar{\tau}$ \\
\hline\hline
BP1 & 26.6 & 0.262 & 0.140 & 13.4 & 0.756 & 46.1 & 10.1 & 2.39\\
\hline
BP2 & 99.999 & - & $<10^{-7}$& $<10^{-7}$  & - & $<10^{-3}$ & $<10^{-3}$ & $<10^{-4}$  \\
\hline
BP3 & - & - & 9.42$\times 10^{-5}$ & - & - & 95.1 & - & 4.93 \\
\hline
BP4 & - & - & 6.65$\times 10^{-8}$ & - & - & 95.1 & - & 4.94 \\
\hline
\end{tabular}
\caption{Decay branching fraction of $A_1$ for the benchmark points where $A_1$ mass calculated at tree level. 
The kinematically forbidden decays are marked with dashes. }\label{A1dcy2}
\end{center}
\end{table}

Next we consider the decay branching fractions of the pseudo-scalar boson, $A_1$ as given in Table~\ref{A1dcy2}.  In the case of BP1, $A_1$ is mostly of doublet type and decays preferably to the fermionic final states, i.e., $b\bar{b}$, $t\bar{t}$ or $\tau\bar{\tau}$. In case of BP2 it is almost a triplet so that the $A_1$ dominantly decays to $W^\pm h^\mp_1$. 
In BP3 and BP4  $A_1$ is of triplet type, but due to the non-available phase space, it decays only into $b\bar{b}$ and $\tau\bar{\tau}$.

\begin{table}
\begin{center}
\renewcommand{\arraystretch}{1.4}
\begin{tabular}{||c|c|c|c|c||}
\hline\hline
$h_1^\pm$ Decay&\multicolumn{4}{|c|}{Branching fractions (\%)}\\
\hline
Modes&$Z W^\pm$ & $h_1 W^\pm$ & $t\bar{b}$ & $\tau \bar{\nu_\tau}$ \\
\hline\hline
BP1 & 56.3 & - & 17.4& 26.3 \\
\hline
BP2 & 41.4 & 9.42 & 49.1 & 3.27$\times 10^{-2}$\\
\hline
BP3 & 59.4 & - & 40.2 & 0.44 \\
\hline
BP4 & 38.7 & 9.72$\times 10^{-2}$ & 12.8 & 48.4  \\
\hline
\end{tabular}
\caption{Decay branching fraction of $h_1^\pm$ for the benchmark points.}\label{ch1dcy}
\end{center}
\end{table}

In Table~\ref{ch1dcy}  we show the decay branching fractions for the charged Higgs boson for each benchmark point. As seen in section~\ref{chdcy}, doublet content $\lsim 1\%$ is enough to open the $tb$ or $\tau \nu$ decay modes. In all four benchmark points we have substantial 
$ZW^\pm$ as well as fermonic decay branching fractions. In the case of BP2 the decay mode $h_1W^\pm$ becomes kinematically allowed.

\begin{table}
\begin{center}
\renewcommand{\arraystretch}{1.4}
\begin{tabular}{||c|c|c|c|c|c|c|c|c||}
\hline\hline

Benchcmark & $\sigma_{h_1^{\pm}tb}$& $\sigma_{h_1^{\pm}t}$&$\sigma_{h_1^{\pm}h_1}$ & $\sigma_{h_1^{\pm}h_2}$ & $\sigma_{h_1^{\pm}A_1}$ & $\sigma_{h_1^{\pm}h_1^{\mp}}$ & $\sigma_{h_1^{\pm} W^{\mp}}$ & $\sigma_{h_1^{\pm} Z}$\\
Points  & fb & fb &fb& fb&fb& fb& fb &fb \\

\hline\hline
BP1 & 11.12 & 16.39 & 2.02 & 301.84 & 3$\times 10^{-2}$ & 151.9 & 225.504 & 0.64 \\
\hline
BP2 & 3.8 & 5.4 &0.39 & 162.1 & 3.9$\times 10^{-2}$ & 80.77 & 1.013 & 0.36 \\
\hline
BP3 &$6.4\times 10^{-3}$ &$8\times10^{-3}$&3$\times 10^{-4}$ & 5$\times 10^{-2}$ & 226.6 & 99.4 & 2.6$\times 10^{-2}$ & 0.27 \\
\hline
BP4 &$1.6\times10^{-3}$ &$3\times10^{-2}$&11$\times 10^{-4}$ & 0.246 & 296.9 & 149.9 & 17.4 & 2.6$\times 10^{-3}$\\
\hline
\end{tabular}
\caption{The important production cross-sections at NLO of four benchmark points 
at the LHC with ECM=14 TeV. The $K$ factor is taken as $K$=1.6.}\label{crossbp}
\end{center}
\end{table}

Table~\ref{crossbp} presents the  cross-sections of the lightest charged Higgs production processes at the LHC. To calculate these cross-sections we obtained related vertices with proper doublet-triplet mixing from {\tt SARAH} \cite{sarah1,sarah2}. The vertex information and mass spectrum are then fed to {\tt CalcHEP} \cite{calchep} to calculate the production cross-sections with 14 TeV center of mass energy at the LHC where we use {\tt CTEQ6L} parton distribution function (PDF) \cite{Lai:1999wy,Pumplin:2002vw}. The renormalization/factorization scale in {\tt CalcHEP} is set at $\sqrt{\hat{s}}$. A $K$-factor of 1.6 is taken for the NLO cross-section evolution \cite{Djouadi:2005gj}. In the numerical analysis we see that the cross-section for $h^+_1h^-_1$ is substantial for all our benchmark points. The $h^\pm_1 h_2$ production cross-section is large for BP1 and BP2 where $h_2$ is degenerate in mass with $h_1^{\pm}$. The production cross-section of $h^\pm_1 A_1$ is large for BP3 and BP4 due to the light $A_1$ while the production cross-section for $h^\pm_1W^\mp$ is large only for BP1 and BP4 since doublet components in $h_2$ and $A_1$ (given in Table \ref{bps}) enhance the contributions coming from the diagrams given in Fig. \ref{h+main_pro}d-e. We also see that due to the large triplet part in the light $h^\pm_1$ the production processes, i.e., $h^\pm_1 t$  and $h^\pm_1 tb$ fail to contribute as much as the other production processes in TESSM. Finally we observe that the production cross-sections of  $h^\pm_1 Z$ and $h^\pm_1 h_1$ are negligible for our BP's and those cross-sections are not taken into account in the collider analysis.
%

\section{Collider simulation}
\label{collidersimulation}

In this study, {\tt SARAH} \cite{sarah1,sarah2} is used to generate the vertices for TESSM. The one-loop lightest neutral Higgs mass spectrum is generated by our self developed mathematica code. The vertex information and mass spectrum are then fed to {\tt CalcHEP} \cite{calchep} which generates the decay branching fractions, which we write in SLHA \cite{slha} format readable to PYTHIA. The events for the simulation were either generated by {\tt PYTHIA (version 6.4.5)} \cite{pythia} or {\tt CalcHEP} \cite{calchep} and later fed to {\tt PYTHIA} for simulation.

For hadronic level simulation we have used {\tt Fastjet-3.0.3} \cite{fastjet} algorithm for the jet formation with the following criteria:
\begin{itemize}
  \item The calorimeter coverage is $\rm |\eta| < 4.5$.
 
  \item As we are looking for leptonic final states we demand 
hard jets, $ p_{T,min}^{jet} = 20$ GeV and jets are ordered in $p_{T}$.
  \item Leptons ($\rm \ell=e,~\mu$) are selected with
        $p_T \ge 10$ GeV and $\rm |\eta| \le 2.5$.
  \item No jet should match with a hard lepton in the event.
   \item $\Delta R_{lj}\geq 0.4$ and $\Delta R_{ll}\geq 0.2$. Here $\Delta R_{ij}=\sqrt{\Delta {\phi}_{ij}+\Delta{\eta}_{ij}}$ is the angle between particle i and particle j, where $\Delta {\phi}_{ij}$ is the difference 
   of the azimuthal angle and $\Delta{\eta}_{ij}$ is the difference of the pseudo-rapidities.  
  \item Since efficient identification
of the leptons is crucial for our study, we required, on top of the above set of
cuts, that hadronic activity within a cone of $\Delta R = 0.3$ between two isolated leptons should be $\leq 0.15 p^{\ell}_T$ GeV in the specified cone.
\end{itemize}

\begin{figure}
\begin{center}
\includegraphics[width=0.75\linewidth]{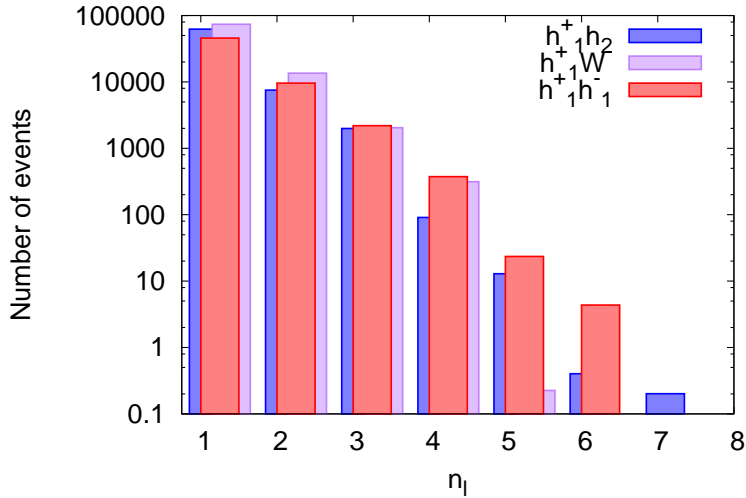}
\caption{ The lepton multiplicity distributions coming from the dominant production processes, $h_1^\pm h_2$, $h_1^\pm W^\mp$ and $h_1^\pm h_1^\mp$, respectively, for
BP1 at the LHC at 1000 fb$^{-1}$ of integrated luminosity.}\label{lmt}
\end{center}
\end{figure}

In Figure~\ref{lmt} we show the lepton multiplicity distribution for
benchmark point 1 (BP1) coming from the dominant production processes, $h^\pm_1 h_2$, $h^\pm_1 W^\mp$ and $h^\pm_1 h^\mp_1$. We see that there are sufficient number of events 
with more than $3\ell$ which can kill many SM backgrounds and reduce substantially the others. In the following sections we demand  $\geq 3\ell$ also use different selection cuts to reduce the SM background further and investigate the minimum required luminosity to have the discovery at $5 \sigma$ signal significance.

\subsection{$3\ell$ final states}
 We can see from the dominant production processes that the charged Higgs, $h^\pm_1$, decays to $ZW^\pm$ can give rise to $3\ell$ final states. The other associated particles like $h_2/A_1$, $W^\pm$ and the other charged Higgs $h^\pm_1$ (in case of the pair production) can contribute in the 
jet-final states through their hadronic decays. Keeping this in our mind and to kill the standard model backgrounds (SM) we look for $\geq 3\ell + (\geq 2-\rm{jet})$ final state. Out of these 3 leptons one pair is coming from the $Z$, so we also demand $|M_{\ell \ell}-M_Z|\leq 5 \, \rm{GeV}$. We choose our final state as $[\geq 3\ell \, (|M_{\ell \ell}-M_Z|\leq 5 \, \rm{GeV})] +\geq 2j  + (\ptmiss \geq 30 \, \rm{GeV})$ and the respective number of events for the benchmark points and the dominant SM backgrounds are listed in 
Table~\ref{3l2j} at an integrated luminosity of 1000 fb$^{-1}$ at the 14 TeV LHC. For the SM backgrounds we have considered productions of $t\bar{t}$, $t\bar{t}Z$, 
$t\bar{t}W^\pm$ and $VV$ (which includes $W^+W^-$, $W^\pm Z$ and $ZZ$) of which $t\bar{t}$ and $VV$ are the dominant ones.  As the numbers are given for an integrated luminosity of 1000 fb$^{-1}$, all four benchmark points cross $5\sigma$ of signal significance; the highest is $\sim 19 \sigma$ for BP1 and lowest is $\sim 9 \sigma$ for BP4.  The earliest $5\sigma$ signal significance comes at an integrated luminosity of  72 fb$^{-1}$  for BP1. For BP4 we have to wait for around 300 fb$^{-1}$ of integrated luminosity. 
\renewcommand{\arraystretch}{1.0}
\begin{table}[bht]
\begin{center}
\begin{tabular}{||c||c|c|c|c||c|c|c|c||}
\hline
\hline
Production &\multicolumn{4}{|c|}{ Benchmark Points }&\multicolumn{4}{|c|}{ Backgrounds }\\
processes &BP1&BP2&BP3&BP4&$t\bar{t}$&$t\bar{t}Z$&$t\bar{t}W$&$VV$\\
\hline
$h^+_1h^-_1$ &809.72&469.11&659.09&480.10&\multirow{5}{*}{2084.34}&\multirow{5}{*}{333.72}&\multirow{5}{*}{190.66}&\multirow{5}{*}{11043.99}\\
$h^\pm_1 W^\mp$ &645.55&0.00&0.00&25.97&&&&\\
$h^\pm_1 \phi$ &903.04&834.23&716.21&581.56&&&&\\
$h^\pm_1 t$ &58.92&18.44&0.00&0.00&&&&\\
$h^\pm_1 tb$ &53.23&17.54&0.00&0.00&&&&\\
 \hline
Total &2470.46&1339.31&1375.30&1087.63&\multicolumn{4}{|c|}{13652.71}\\
\hline
Significance &18.64&10.66&11.22&8.96&\multicolumn{4}{|c|}{}\\
\hline
\hline
\end{tabular}
 \caption{Number of events after the selection cuts for $\geq 3\ell \, (|M_{\ell \ell}-M_Z|\leq 5 \, {\rm GeV}) +\geq 2j  + (\ptmiss \geq 30 \, \rm{GeV})$ final state for the benchmark points and backgrounds at an integrated luminosity of 1000 fb$^{-1}$ at the LHC with $E_{cm}=14$ TeV.  Here $h^\pm_1 \phi$ includes contributions from both $h^\pm_1 h_2$ and $h^\pm_1 A_1$.}
\label{3l2j}
\end{center}
\end{table}

Next we focus on the possibility to tag the charged Higgs boson in association with a neutral Higgs, i.e., $h_2$ or/and $A_1$. We can see from Table~\ref{H1dcy2} and Table~\ref{A1dcy2} that either of these two Higgses decays to $b\bar{b}$ substantially except in BP2. In the case of BP1 and BP2 it is mainly $h^\pm_1h_2$ that contributes due to lighter masses of $h_2$  and higher production rates (see Table~\ref{crossbp}). The situation is quite opposite in the case of BP3 and BP4 where $h^\pm_1A_1$ contributes instead of $h^\pm_1h_2$. 

For the final state we demand $\geq 3\ell \, (|M_{\ell \ell}-M_Z|\leq 5\, {\rm GeV}) +(\geq 2b-{\rm jets})  +( \ptmiss \geq 30 \, \rm{GeV})$, where we expect the two $b$ jets mostly come from the neutral Higgses. The charged 
Higgs pair production $h^+_1h^-_1$ also  contributes to the final state when one of the charged Higgs in the pair production decays to $ZW^{\pm}$ and the other decays to $tb$ which is one of the substantial decay channels especially for BP2 and BP3 with Br($h^\pm_1 \to tb\gsim 40\%$) (see Table~\ref{ch1dcy}). The aforementioned  final state can be obtained even when both the charged Higgs bosons decay to $ZW^{\pm}$, but this time one of the $Z$ boson decays to $b\bar{b}$. Table~\ref{3l2b} gives the number of events in benchmark points for the signal and the SM backgrounds 
at an integrated luminosity of 1000 fb$^{-1}$ at the LHC with $E_{cm}=14$ TeV. As expected $h^\pm_1 \phi$ contributes most. In the case of BP2 the contribution is much smaller than the Br($h_2 \to b\bar{b})\sim 1\%$ (see Table~\ref{H1dcy2}). The next largest contribution comes from the $h^\pm_1$ pair production as explained above.  The signal significances can be read from Table~\ref{3l2b} and again BP1 is the first to give the $5\sigma$ significance at an integrated luminosity of $\sim 100 \, \rm{fb}^{-1}$. For BP2 it takes around 1000 fb$^{-1}$ of integrated luminosity to reach the $5\sigma$ significance. 
\renewcommand{\arraystretch}{1.0}
\begin{table}
\begin{center}
\begin{tabular}{||c||c|c|c|c||c|c|c|c||}
\hline
\hline
Production &\multicolumn{4}{|c|}{ Benchmark Points }&\multicolumn{4}{|c|}{ Backgrounds }\\
processes &BP1&BP2&BP3&BP4&$t\bar{t}$&$t\bar{t}Z$&$t\bar{t}W$&$VV$\\
\hline
$h^+_1h^-_1$ & 30.07&59.93&54.39&13.34&\multirow{5}{*}{99.25}&\multirow{5}{*}{143.78}&\multirow{5}{*}{69.03}&\multirow{5}{*}{118.75}\\
$h^\pm_1 W^\mp$ &30.44&0.00&0.00&0.81&&&&\\
$h^\pm_1 \phi$ &410.27&60.29&358.22&278.76&&&&\\
$h^\pm_1 t$ &4.64&1.57&0.00&0.00&&&&\\
$h^\pm_1 tb$ &9.66&3.08&0.00&0.00&&&&\\
 \hline
Total &485.08&124.87&412.61&292.91&\multicolumn{4}{|c|}{430.81}\\
\hline
Significance &15.68&5.12&14.21&10.89&\multicolumn{4}{|c|}{}\\
\hline
\hline
\end{tabular}
 \caption{Number of events after the selection cuts for $\geq 3\ell \, (|M_{\ell \ell}-M_Z|\leq 5 {\rm GeV}) +(\geq 2b-{\rm jets})  + {\ptmiss \geq 30 \, \rm{GeV}}$  final state for the benchmark points and backgrounds at an integrated luminosity of 1000 fb$^{-1}$ at the LHC with $E_{cm}=14$ TeV.  Here $h^\pm_1 \phi$ includes contributions from both $h^\pm_1 h_2$ and $h^\pm_1 A_1$.}
\label{3l2b}
\end{center}
\end{table}

From Table~\ref{ch1dcy} we can see that for BP1 and BP4 the charged Higgs also decays to $\tau \nu$ significantly.  Also $W^\pm$ can decay to $\tau\nu$, and thus $h^\pm_1W^\mp$ and the other production processes 
contribute to the single tau final states. In Table~\ref{3l1tau} we have considered $\geq 3\ell \, (|M_{\ell \ell}-M_Z|\leq 5 {\rm GeV}) + (\geq 1\tau-{\rm jet}) + (\ptmiss \geq 30 \, \rm{GeV})$.  The demand of an extra tau-jet on top of three leptons ($e$ and/or $\mu$) reduces the backgrounds substantially. Here by $\tau$-jet we considered the hadronic decay of the $\tau$ with at least one charged track within $\Delta R \leq 0.1$ of the candidate $\tau$-jet \cite{taujet}. Table~\ref{3l1tau} presents the corresponding numbers at 1000 fb$^{-1}$ of integrated luminosity at the 14 TeV LHC.  We can see that except BP3 all the other benchmark points give $> 5 \sigma$. The earliest discovery reach is for BP1 and it takes around 146 fb$^{-1}$ of integrated luminosity.

One can also look for $\geq 3\ell  + (\geq 2\tau-\rm{jet}) $, where the tau pair is coming from the decays 
of the neutral Higgses ($h_2, A_1$).  We have analysed this particular final state as well and it turns out
to require $\gsim 1000$ fb$^{-1}$ integrated luminosity for a discovery.

\renewcommand{\arraystretch}{1.0}
\begin{table}[bht]
\begin{center}
\begin{tabular}{||c||c|c|c|c||c|c|c|c||}
\hline
\hline
Production &\multicolumn{4}{|c|}{ Benchmark Points }&\multicolumn{4}{|c|}{ Backgrounds }\\
processes &BP1&BP2&BP3&BP4&$t\bar{t}$&$t\bar{t}Z$&$t\bar{t}W$&$VV$\\
\hline
$h^+_1 h^-_1 $ & 221.00&36.10&50.52&225.58&\multirow{5}{*}{0.00}&\multirow{5}{*}{25.06}&\multirow{5}{*}{0.00}&\multirow{5}{*}{237.51}\\
$h^\pm_1  W^\mp$ &55.47&0.00&0.00&1.91&&&&\\
$h^\pm_1  \phi$ &37.72&70.18&25.38&19.00&&&&\\
$h^\pm_1  t$ &3.80&1.07&0.00&0.00&&&&\\
$h^\pm_1  tb$ &2.96&0.05&0.00&0.00&&&&\\
\hline
Total &320.96&108.41&75.89&246.50&\multicolumn{4}{|c|}{262.57}\\
\hline
Significance &13.08&5.53&4.13&10.93&\multicolumn{4}{|c|}{}\\
\hline
\hline
\end{tabular}
 \caption{Number of events  for $\geq 3\ell \, (|M_{\ell \ell}-M_Z|\leq 5 {\rm GeV}) +(\geq 1\tau-jet)  + (\ptmiss \geq 30 \, \rm{GeV})$ final state for the benchmark points and backgrounds at an integrated luminosity of 1000 fb$^{-1}$ at the LHC with $E_{cm}=14$ TeV.  Here $h^\pm_1 \phi$ includes contributions from both $h^\pm_1 h_2$ and $h^\pm_1 A_1$.}
\label{3l1tau}
\end{center}
\end{table}
 
\subsection{$4\ell$ final states}

We explore the possibility where the associated particle to the light charged Higgs boson $h^\pm_1$
can contribute to the leptonic final states. Thus we expect all the production processes to contribute in $\geq 4 \ell$ the final states. The number of events for $\geq 4\ell + (\ptmiss \geq 30 \, \rm{GeV})$ at an integrated luminosity of 1000 fb$^{-1}$ at the LHC with $E_{cm}=14$ TeV are listed in Table~\ref{4l}. $t\bar{t}$ and $VV$ still contribute as dominant backgrounds. At 1000 fb$^{-1}$ BP1 and BP2 give 11 $\sigma$ and 9 $\sigma$, respectively, whereas BP3 and BP4 give nearly $5 \sigma$ significance. 
\renewcommand{\arraystretch}{1.0}
\begin{table}[bht]
\begin{center}
\begin{tabular}{||c||c|c|c|c||c|c|c|c||}
\hline
Production &\multicolumn{4}{|c|}{ Benchmark Points }&\multicolumn{4}{|c|}{ Backgrounds }\\
processes &BP1&BP2&BP3&BP4&$t\bar{t}$&$t\bar{t}Z$&$t\bar{t}W$&$VV$\\
\hline
$h^+_1 h^-_1 $ &344.82&152.82&223.34&243.57&\multirow{5}{*}{1191.04}&\multirow{5}{*}{651.62}&\multirow{5}{*}{11.5}&\multirow{5}{*}{1425.03}\\
$h^\pm_1  W^\mp$ &283.65&0.00&0.00&12.70&&&&\\
$h^\pm_1  \phi$ &85.73&407.63&41.92&27.01&&&&\\
$h^\pm_1  t$ &14.80&4.54&0.00&0.00&&&&\\
$h^\pm_1  tb$ &10.00&3.17&0.00&0.00&&&&\\
\hline
Total &739.00&568.15&265.26&283.28&\multicolumn{4}{|c|}{3279.19}\\
\hline
Significance &11.30&9.04&4.46&4.75&\multicolumn{4}{|c|}{}\\
\hline
\hline
\end{tabular}
 \caption{Number of events after the selection cuts for $\geq 4\ell + (\ptmiss \geq 30 \, \rm{GeV})$  final states for the benchmark points and backgrounds at an integrated luminosity of 1000 fb$^{-1}$ at the LHC with $E_{cm}=14$ TeV.   Here $h^\pm_1 \phi$ includes contributions from both $h^\pm_1 h_2$ and $h^\pm_1 A_1$.}
\label{4l}
\end{center}
\end{table}

\subsection{$5\ell$ final states}

If both the charged Higgses in the pair production decay to $ZW^\pm$, then we can get $5\ell$ and $6\ell$ final sates depending on the leptonic decays of the $Z$ and $W^\pm$.  Associated neutral Higgs production $h^\pm_1\phi$ can also contribute to $5\ell$ final sates if the neutral Higgses can decay in the leptonic modes. In Table~\ref{5l} we present the numbers of events for the $\geq 5\ell + (\ptmiss \geq 30 \, \rm{GeV})$ final state. We see that due to large  Br($h_2\to ZZ)\sim 44\%$, $h^\pm_1\phi$
contributes more to $5\ell$ final state. Thus BP2 makes to $\gsim 9 \sigma$ at 1000 fb$^{-1}$ and BP1 crosses $5\sigma$ in the signal significance.  

In addition, we have analysed the final state with two more jets ($\geq 5\ell + (2-{\rm jet})+ (\ptmiss \geq 30 \, \rm{GeV})$) which comes from the hadronic decay of the second $W^\pm$ of the second charged Higgs in the case of charged Higgs pair production. In this case only charged Higgs pair production contributes to the final sate. Although this case is almost background free, the number of signal events is also quite small. Thus only for BP2 it makes to  $\sim 5\sigma$ even at 1000 fb$^{-1}$ and for the rest of the benchmark points it either cross $3\sigma$ or approach to $3\sigma$.  When the second $W^\pm$ also decays leptonically we get $6\ell +\ptmiss$ in the final state. This final state is absolutely background free
but with low signal event counts, and only BP1 and BP2 can cross $2\sigma$ signal significance.

\renewcommand{\arraystretch}{1.0}
\begin{table}[bht]
\begin{center}
\begin{tabular}{||c||c|c|c|c||c|c|c|c||}
\hline
Production &\multicolumn{4}{|c|}{ Benchmark Points }&\multicolumn{4}{|c|}{ Backgrounds }\\
processes &BP1&BP2&BP3&BP4&$t\bar{t}$&$t\bar{t}Z$&$t\bar{t}W$&$VV$\\
\hline
$h^+_1 h^-_1 $ &22.78&14.62&21.48&13.19&\multirow{5}{*}{0.00}&\multirow{5}{*}{2.64}&\multirow{5}{*}{1.64}&\multirow{5}{*}{0.00}\\
$h^\pm_1  W^\mp$ &0.23&0.00&0.00&0.00&&&&\\
$h^\pm_1  \phi$ &11.30&75.04&3.39&2.97&&&&\\
$h^\pm_1  t$ &0.03&0.00&0.00&0.00&&&&\\
$h^\pm_1  tb$ &0.04&0.004&0.00&0.00&&&&\\
\hline
Total &34.38&89.67&24.87&16.16&\multicolumn{4}{|c|}{4.28}\\
\hline
Significance &5.52&9.25&4.61&3.57&\multicolumn{4}{|c|}{}\\
\hline
\hline
\end{tabular}
 \caption{Number of events after the selection cuts for $\geq 5\ell + (\ptmiss \geq 30 \, \rm{GeV})$  final state for the benchmark points and backgrounds at an integrated luminosity of 1000 fb$^{-1}$ at the LHC with $E_{cm}=14$ TeV.   Here $h^\pm_1 \phi$ includes contributions from both $h^\pm_1 h_2$ and $h^\pm_1 A_1$.}
\label{5l}
\end{center}
\end{table}

\section{Mass distributions}
\label{discussre}

\begin{figure}[thb]
\begin{center}
\includegraphics[width=0.45\linewidth]{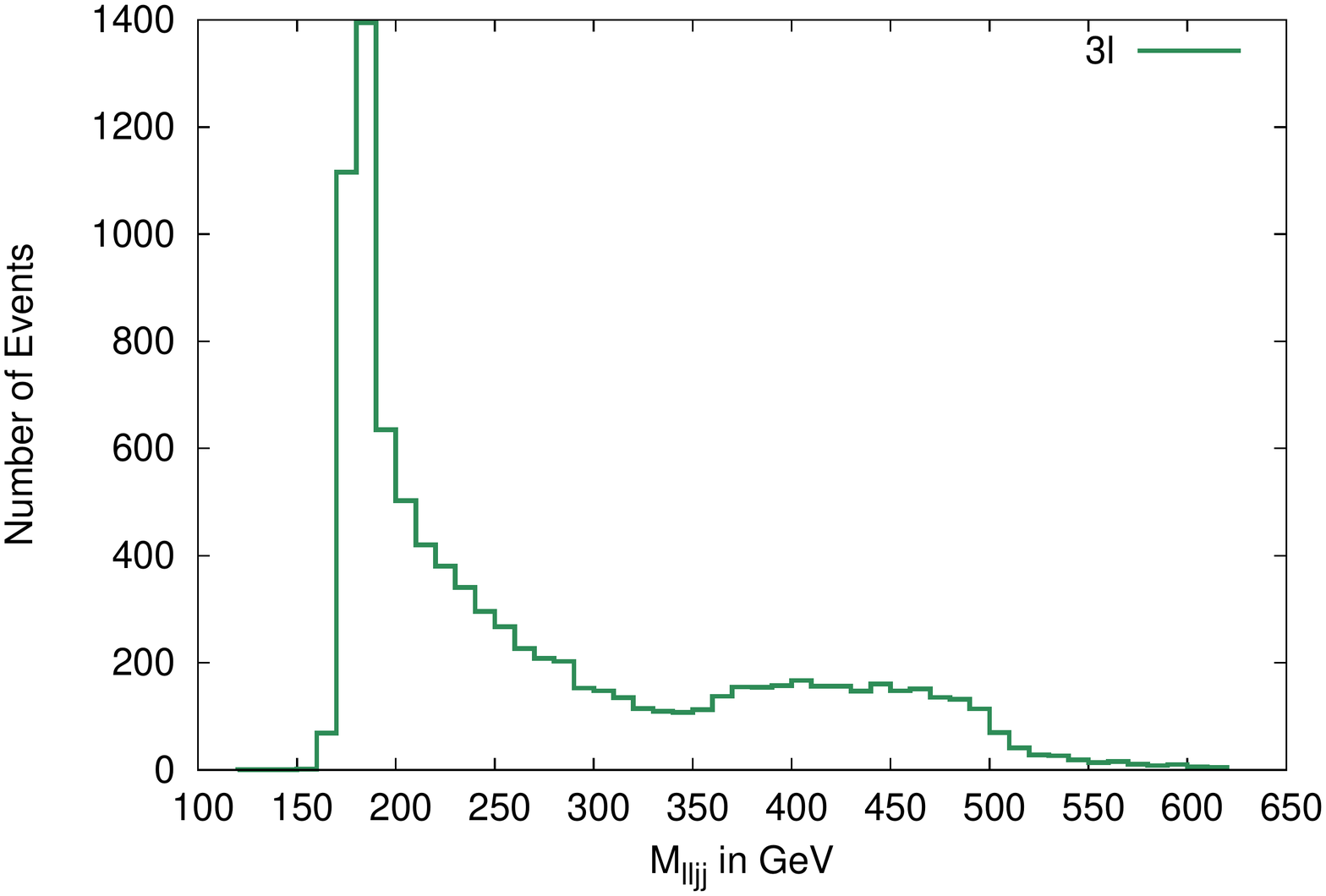}
\includegraphics[width=0.45\linewidth]{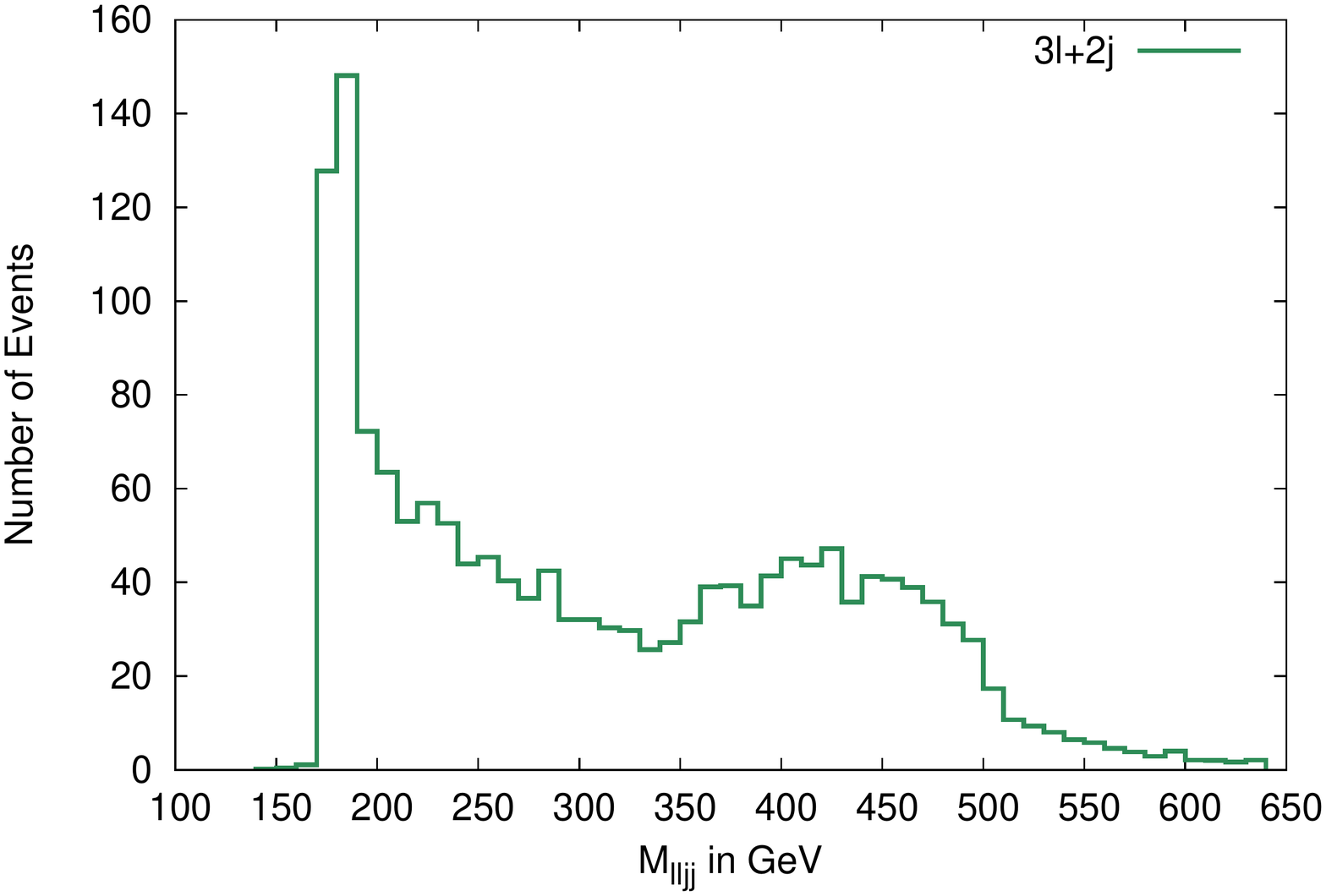}
\caption{ The $\ell\ell jj$ invariant mass distributions coming from the signal for BP1 for $\geq 3\ell+(\ptmiss \geq 30$ GeV) (left) and $\geq 3\ell+2j+(\ptmiss \geq 30$ GeV) (right) final states at an integrated luminosity of 1000 fb$^{-1}$.}\label{2l2jinv}
\end{center}
\end{figure}

When the charged Higgs $h^\pm_1$ decays to $ZW^\pm$, it is possible to reconstruct the mass of the 
charged Higgs via constructing the invariant mass of di-lepton and di-jet; where the di-lepton comes from $Z$ and the di-jet from the $W^\pm$. If we demand that the associated particle ($W^\pm$, $h_2$, $A_1$ or $h^\pm_1$) provides the other lepton such that the final state has $\geq 3\ell$, it can kill the backgrounds substantially.  In Figure~\ref{2l2jinv} (left) we present the invariant mass distribution of $M_{\ell\ell jj}$ for BP1 with $(\geq 3\ell )+(\ptmiss \geq 30$ GeV) final state, where $| M_{\ell\ell}-M_Z| \leq 5 \, \rm{GeV}$ and $| M_{jj}-M_{W^\pm}| \leq 10 \, \rm{GeV}$ cuts are applied to ensure the reconstruction of $Z$ and $W^\pm$, respectively. Whenever the $Z$ and $W^\pm$ come from the decay of the light charged Higgs $h^\pm_1$, i.e. with the right combination, we get the invariant mass peaked around $M_{h^\pm_1}$. The wrong combination will give us the threshold of the production process (in case of $h^\pm_1W^\mp$) which can be seen as the edge after the hump in the distribution. In Figure~\ref{2l2jinv} (right) we present the invariant mass distribution of $M_{\ell\ell jj}$ for BP1 with $(\geq 3\ell )+\geq 2j+(\ptmiss \geq 30$ GeV) final state. Although the number of events decreases compared to the previous one, the end edge distribution is much clearer now. 
In case of $(\geq 3\ell )+\geq 2j+(\ptmiss \geq 30$ GeV) (Figure~\ref{2l2jinv} (left)), Higgs mass peak around 183 GeV gets $7.6\sigma$ signal significance over SM backgrounds whereas for $(\geq 3\ell )+\geq 2j+(\ptmiss \geq 30$ GeV) (Figure~\ref{2l2jinv} (left)), the significance is around $7.3\sigma$.

\begin{figure}[bht]
\begin{center}
\includegraphics[width=0.75\linewidth]{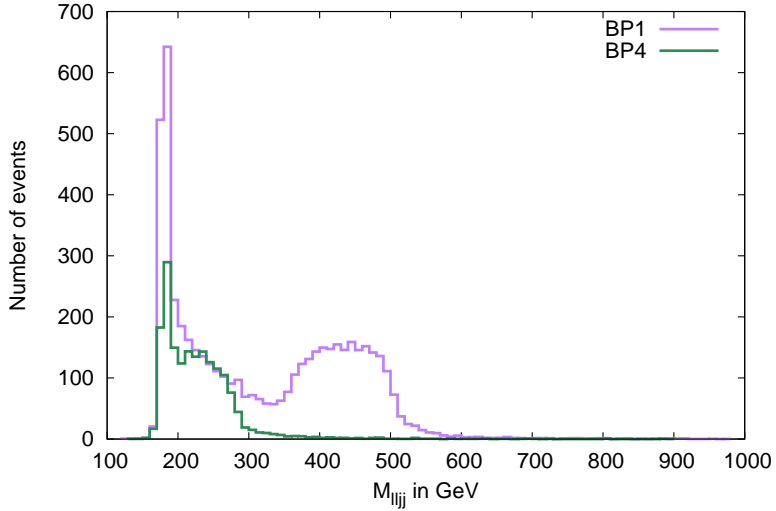}
\caption{ $h^\pm_1W^\pm$ contribution to  $\ell\ell jj$ invariant mass distribution
for $\geq 3\ell+(\ptmiss \geq 30$ GeV ) final state for BP1 and BP4 
at an integrated luminosity of 1000 fb$^{-1}$. In BP4 number of events are multiplied by 5.}\label{2l2j}
\end{center}
\end{figure}

In particular,  Figure~\ref{2l2j} shows the corresponding contribution of $h^\pm_1 W^\mp$ production channel for BP1 and BP4.  Clearly we can see the charged Higgs mass peak around 180 GeV but there is an edge at the end of the invariant mass distribution, which is the artifact of the s-channel threshold i.e., $h_2$ or $A_1$. In other words, the edge at different mass values for different benchmark points carries hidden information about the heavier Higgs bosons. For BP1 the edge is around $\gsim 500$ GeV, whereas for BP4 it comes around  $\gsim 300$ GeV (see Figure~\ref{2l2j}). In the case of BP1 the dominant contribution comes from $A_1$ as it is most doublet type (see Table~\ref{bps}) neutral Higgs so that it dominantly couples to the quarks. In the case of BP4, it is the $h_2$ which has relatively more doublet component, so dominantly contribute in the production process. 

Figure~\ref{2l2jinv} shows that all the dominant production processes
give a peak around the right charged Higgs mass.  The demand of $\geq 3$ lepton final state
makes the distribution almost background free. So these multi-lepton final sates not only 
probe $Z-W^\pm-h^\mp_i$ coupling but also get the mass information of the light charged Higgs as well as the heavier neutral Higgs boson. The non-standard decays of charged Higgs opens a new era 
of light charged Higgs phenomenology and the multi-lepton final states along with the mass reconstruction could be a very handy tools to probe these decays. 

Above mentioned multi-lepton signatures carry the information of the charged Higgs decay to $ZW^\pm$  and the discovery of such a final state can provide the information about the presence of higher multiplet structure in the Higgs sector.  
Besides the production channels containing non supersymmetric particles, the charged Higgs productions via supersymmetric decays are quite interesting and some studies have already been performed in the context of MSSM \cite{pbch}. We are also working on similar analysis
which would also provide some information about the structure of the Higgs sector as well as the supersymmetric electroweak gauginos and the third generation squark mass spectrum \cite{triplino}. 

\section{Conclusions}
\label{conclusionsc}
In this article we have studied the multi-lepton signatures of triplet like charged Higgs at the LHC  in the $Y=0$ triplet extended supersymmetric model (TESSM). First of all, to understand the triplet impact on charged Higgs sector of the model we performed a parameter scan where we calculated the lightest neutral Higgs boson mass numerically at one-loop level and we collected points satisfying the direct search constraints along with the measured Higgs mass. We have shown that for many points, the lightest charged Higgs is non-degenerate in mass with  CP odd and heavy CP even Higgs bosons. We observed that the presence of the non-degeneracy allows on-shell CP odd or CP even Higgs boson to decay into a charged Higgs and $W^{\pm}$ ($h_2(A_1)\rightarrow h_1^{\pm} W^{\mp}$) which is not kinematically possible in the case of the MSSM.  

In addition to the non-degeneracy, the existence of a triplet with non-zero vev provides the tree level $h_i^{\pm} Z W^{\mp}$ coupling that is only induced at loop order in the case of two Higgs doublet models. This vertex can be used to discriminate the triplet nature among the other structures in the Higgs sector. For this purpose we have investigated the strength of the coupling for different scenarios with the charged Higgs mass $m_{h_1^{\pm}}\leq 200$ GeV and shown that for triplet like charged Higgs, the $ZW^{\pm}$ decay channel can be as substantial as $tb$ and $\tau \nu_{\tau}$. In particular, we have shown that when the lightest charged Higgs comprises $99 \%$ triplet nature the $ZW^{\pm}$ channel remains the most dominant channel even after $tb$ channel is kinematically open. 

Besides the decay channels, we have investigated the production channels for the triplet like charged Higgs at the LHC to distinguish the signatures of the model from the ones in the two Higgs doublet models. Due to the lack of triplet-SM fermion coupling, the production channel through the top decay $t\rightarrow b h_1^{\pm}$ or the associated production with a top quark through the gg/gb fusions are no longer the dominant ones for the triplet like charged Higgs.  This is why we have explored alternative production channels such as charged Higgs pair production, associated production with massive gauge bosons ($W^{\pm}, Z$) and neutral Higgs bosons. We have also considered the charged Higgs production through vector boson fusion that is possible with the presence of $h_i^{\pm} Z W^{\mp}$ coupling.  Then we discussed the possibility of having triplet like neutral Higgs boson along with the charged ones where the production mechanisms through the neutral Higgs propagators are expected to be suppressed. 

To determine the dominant production channels we have considered four benchmark points where we performed PYTHIA level simulation using FastJet jet information at the LHC with 14 TeV.  We analysed the signal in $3l,4l,5l$ final states with $\tau$ and b-jets. It is possible to reduce the SM background substantially by using the different selection cuts for the multi-lepton and jets and the earliest discovery hints can be obtained with $\gtrsim 72$ fb$^{-1}$ integrated luminosity.  We have also presented the invariant mass distribution $M_{lljj}$ for ($\geq 3\ell)+(\ptmiss \geq 30$ GeV) and $(\geq 3\ell)+(\geq 2j)+(\ptmiss \geq 30$ GeV) and shown that in addition to the charged Higgs mass peak, an edge that carries information about heavy intermediate neutral Higgs bosons emerges at the end of the mass distribution. 

{\bf Acknowledgments:} 
The authors gratefully acknowledge support from the Academy of Finland (Project No. 137960). The work of ASK is also supported by the Finnish Cultural Foundation. PB also thanks INFN, Lecce, for the support at the finishing part of the project.

\end{document}